\documentclass[
    prd, twocolumn, superscriptaddress, nofootinbib, amsmath, amssymb,
    aps, floatfix, longbibliography
]{revtex4-2}
\usepackage[utf8]{inputenc}
\usepackage{graphicx,xcolor,setspace}
\usepackage[
    colorlinks=true,
    linkcolor=blue,
    urlcolor=blue,
    citecolor=blue
]{hyperref}
\usepackage{slashed,physics}
\usepackage[capitalise]{cleveref}
\usepackage{siunitx}
\usepackage{xspace}
\usepackage{enumerate}

\DeclareSIUnit{\year}{yr}

\usepackage{bm}

\newcommand{\Br}{{\rm BR}}

\newcommand{\fL}{\mathcal{L}}
\newcommand{\fM}{\mathcal{M}}

\newcommand{\LamQCD}{\Lambda_{\rm QCD}}
\newcommand{\sigv}{\left\langle \sigma v\right\rangle}
\newcommand{\TRH}{T_{\rm RH}}
\newcommand{\TQCD}{T_{\rm QCD}}
\newcommand{\mPl}{m_{\rm Pl}}

\newcommand{\sm}{{\rm SM}}
\newcommand{\eV}{{\rm eV}}
\newcommand{\keV}{{\rm keV}}
\newcommand{\MeV}{{\rm MeV}}
\newcommand{\GeV}{{\rm GeV}}
\newcommand{\TeV}{{\rm TeV}}

\newcommand{\no}{\nonumber}
\def\beq{\begin{equation}}
\def\eeq{\end{equation}}
\def\bea{\begin{eqnarray}}
\def\eea{\end{eqnarray}}

\newcommand{\eg}{\textit{e.g.}}

\definecolor{blue-violet}{rgb}{0.54, 0.17, 0.89}

\begin{document}


 \title{Dark Matter Through the Axion-Gluon Portal}

\author{Patrick J. Fitzpatrick}
\affiliation{Racah Institute of Physics, Hebrew University of Jerusalem, Jerusalem 91904, Israel}
\affiliation{Technion---Israel Institute of Technology, Haifa 32000, Israel}

\author{Yonit Hochberg}
\affiliation{Racah Institute of Physics, Hebrew University of Jerusalem, Jerusalem 91904, Israel}

\author{Eric Kuflik}
\affiliation{Racah Institute of Physics, Hebrew University of Jerusalem, Jerusalem 91904, Israel}

\author{Rotem Ovadia}
\affiliation{Racah Institute of Physics, Hebrew University of Jerusalem, Jerusalem 91904, Israel}

\author{Yotam Soreq}
\affiliation{Technion---Israel Institute of Technology,  Haifa 32000, Israel}

\date{\today}

\begin{abstract}
Axion-like-particles are a well-motivated extension of the Standard Model that can mediate interactions between the dark matter and ordinary matter. Here we consider an axion portal between the two sectors, where the axion couples to dark matter and to QCD gluons. We establish the relevant processes of interest across the scales of dark matter and axion masses and couplings, identify the distinct mechanisms that control the dark matter relic abundance in each case, and extract the resulting experimental signatures of the gluonic axion portal to dark matter.   
\end{abstract}

\maketitle

\flushbottom

\section{Introduction}
\label{sec:intro}

Dark matter~(DM) makes up the vast majority of the matter in our universe, but we still know little about its particle nature. 
Many processes can be considered to set its relic abundance in the early universe, including $2\to2$ annihilations~\cite{Lee:1977ua,Griest:1990kh,DAgnolo:2015ujb,DAgnolo:2017dbv,DAgnolo:2019zkf,Kim:2019udq,Frumkin:2022ror}, $n\to2$ annihilations~\cite{Carlson:1992fn,Hochberg:2014dra,Kopp:2016yji,Soni:2016gzf}, as well as decays and inverse decays~\cite{Hall:2009bx,Frumkin:2021zng}. 
(For recent reviews, see \eg{} Refs.~\cite{Battaglieri:2017aum,Asadi:2022njl}.)  
In many frameworks, couplings between the DM and the Standard Model~(SM) should be present in order to mediate interactions between the dark and visible particles, serving as a `portal' between the sectors.

A well-motivated portal is that of the axion, or axion-like-particle~(ALP). 
ALPs as mediators to the dark sector have been studied in  Refs.~\cite{Nomura:2008ru,Freytsis:2010ne,Essig:2013lka,Dolan:2014ska,Kamada:2017tsq,Kaneta:2017wfh,Hochberg:2018rjs,Berlin:2018bsc,deNiverville:2019xsx,Darme:2020sjf,Ge:2021cjz,Gola:2021abm,Domcke:2021yuz,Zhevlakov:2022vio,Bauer:2022rwf,Bharucha:2022lty,Ghosh:2023tyz}.
In this paper, we consider ALP couplings to SM gluons. 
Via this coupling, the ALP essentially couples to the QCD states of the SM, whose degrees of freedom change as one flows through the QCD confining phase transition in the early universe. 
Throughout the cosmological history, one can study the relative importance of the various processes that can occur, and determine their impact on the DM relic abundance.
The DM abundance can result from either freeze-out or freeze-in processes, with a variety of existing experiments placing important constraints and many future experiments set to probe novel regions of the parameter space. 

This paper is organized as follows. 
In Section~\ref{sec:model}, we introduce the model and in Section~\ref{sec:ADMprod} we study the axion and DM relic abundance. 
Sections~\ref{sec:FI} and~\ref{sec:FO} analyze the DM phases of the theory, which include freeze-in and freeze-out, respectively. Current experimental constraints and future probes are presented in Section~\ref{sec:exp}. 
We present our results in Section~\ref{sec:results} and conclude in Section~\ref{sec:conc}.  
Appendix~\ref{app:rates} presents the thermally averaged rate calculations. 
In Appendix~\ref{app:freeze-in}, we elaborate on the analytical estimates for freeze-in. 
Appendix~\ref{app:model} adds further details about the model.  
Appendix~\ref{app:casting constraints} explains in detail the experimental bounds presented in this work. 
%

\section{Model}
\label{sec:model}

We begin by presenting the theory that we will consider in this work. 
The model is an extension of the SM with an axion, $a$, and a Dirac fermion DM candidate, $\chi$. 
Generalizing to a Majorana fermion is straightforward.
We consider the axion, at some UV scale, to couple only to gluons and to the DM. 
This is similar to a KSVZ axion~\cite{Kim:1979if, Shifman:1979if} where the heavy quarks are electroweak singlets.

The effective Lagrangian at the UV scale $\Lambda = 8 \pi f_a$ is given by
\begin{align}
    \label{eq:lagrangian}
	\fL
=&  \fL_{\rm SM}+\frac{1}{2} \partial^\mu a \partial_\mu a 
    - \frac{m_a^2}{2} a^2 + i \overline{\chi} \slashed{D} \chi 
    -  m_\chi \overline{\chi} \chi \nonumber \\
&   - i c_\chi m_\chi \frac{ a }{f_a} \overline{\chi} \gamma_5 \chi  
    - \frac{\alpha_s}{8 \pi } \frac{ a }{f_a} G^{a \mu\nu} \tilde{G}_{a \mu \nu}\,.
\end{align}
Here $G^{a \mu\nu}$ is the gluon field strength, ${\tilde{G}^a_{\mu\nu} \equiv \frac{1}{2} \epsilon_{\mu\nu \rho \sigma} G^{a \rho \sigma}}$ is its dual, $\alpha_s$ is the coupling strength of the strong force, $m_a$ and $m_\chi$ are the axion and DM masses, respectively, $f_a$ is the axion decay constant defined by its coupling to gluons as above, and $c_\chi$ is a dimensionless coefficient parameterizing the coupling of the axion to the DM relative to its coupling to the SM gluons.  
For simplicity, we take $c_\chi=1$ throughout this work. 
The RG flow will create loop-induced couplings to other SM particles at lower scales as one flows below $\Lambda$. 
We limit ourselves to cases where $m_\chi, m_a < 8 \pi f_a$ for validity of our computations.   
The axion mass term is the sum of a dynamical term (which determines the mass of the QCD axion) $m_{a,{\rm QCD}}^2 \simeq m_\pi^2{f_\pi^2}/{ f_a^2}  $ and a bare term $m_{a,0}^2$. 
To avoid fine-tuning of the contributions to the axion  mass  we consider only $m_a \geq m_{a,{\rm QCD}}$.
Note that an axion that solves the CP problem but is heavier than the standard QCD axion could realize the framework presented here, as in {\it e.g.} Refs.~\cite{Rubakov:1997vp,Fukuda:2015ana,Fukuda:2017ywn,Hook:2019qoh,Dunsky:2023ucb}. 

At temperatures and axion masses well above the QCD confinement scale $\LamQCD \sim 200\,\MeV$, QCD can be treated perturbatively and therefore we need only to take into account axion-gluon interactions. 
However, close to and below the QCD scale, one must consider axion-hadron interactions. 
These interactions have been studied in detail for scales less than $ 4\pi f_\pi$ using chiral perturbation theory~($\chi$PT) in Ref.~\cite{Georgi:1986df}. 
This analysis has been extended to scales in the range of $1-3\,\GeV$ in Ref.~\cite{Aloni:2018vki} by using data driven methods. 
Close to $\LamQCD$, the dynamics of the axion are governed by kinetic and mass mixing with the $\pi^0$, $\eta$, $\eta'$ mesons. At scales much below $\LamQCD$, the leading order dynamics stem from a loop-induced coupling to photons, 
\begin{align}
    \label{eq:aFFtil}
    \mathcal{L} 
    \supset 
    -\frac{c_{\gamma} \alpha_{\rm EM}}{8 \pi} \frac{a}{f_a} F_{\mu\nu} \tilde{F}^{\mu\nu}\,,
\end{align}
with $F_{\mu\nu}$ and  $\tilde F_{\mu\nu}$ the photon field strength and its dual and $\alpha_{\rm EM}$ the electromagnetic coupling strength. 
We take $c_\gamma = 1.92$ at low energy, matching the KSVZ axion~\cite{Georgi:1986df, GrillidiCortona:2015jxo}.

\section{Axion and dark matter production}
\label{sec:ADMprod}

We begin by writing down the Boltzmann equations governing the evolution of the axion and DM abundances. 
The relevant Boltzmann equations are
\begin{align}
    \label{eq:BoltzmannEqs}
    \dot{n}_a+3 H n_a 
=&   -\Gamma_{a\,\sm \to \sm } \left(n_a -n_a^{\rm eq}\right) \nonumber\\
     &- \Gamma_{a \to \chi \bar{\chi} } \left(n_a -n_\chi^2 \frac{n_a^{\rm eq}}{{n_\chi^{\rm eq}}^2}\right) \nonumber\\
     &+ \sigv_{\chi \bar{\chi} \to a a }\left(n_\chi^2 -n_a^2 \frac{{n_\chi^{\rm eq}}^2}{{n_a^{\rm eq}}^2}  \right) \,,
\end{align}
and
\begin{align}
    \label{eq:FI boltzmann eq}
    \dot{n}_\chi+3 H n_\chi 
 =& -\sigv_{ \chi \bar{\chi} \to \sm}^{\rm sub}
    \left(n_\chi^2 -{n_\chi^{\rm eq}}  \right) \nonumber\\
    &- \sigv_{\chi \bar{\chi} \to a a }\left(n_\chi^2 -n_a^2 \frac{{n_\chi^{\rm eq}}^2}{{n_a^{\rm eq}}^2}  \right)\nonumber\\
    &+ \Gamma_{a \to \chi \bar{\chi} } \left(n_a -n_\chi^2 \frac{n_a^{\rm eq}}{{n_\chi^{\rm eq}}^2}\right) \nonumber\\
 &+ \!\! \sum_{P=\pi^0,\eta,\eta'}\!\!\Gamma_{P \to \chi \bar{\chi} } \left(n_P^{\rm eq} -n_\chi^2 \frac{n_P^{\rm eq}}{{n_\chi^{\rm eq}}^2}\right)\,.  
\end{align}
Here $n_i^{\rm eq}$ are the equilibrium abundances for the different species and $H$ is the Hubble parameter.
For meson densities we take the densities to vanish at temperatures above the QCD phase-transition temperature $\TQCD$ and as a Bose-Einstein distribution below it, $n_P^{\rm eq} = n_P^{\rm BE}(T) \Theta(\TQCD-T)$~\cite{Husdal:2016haj}. 
We discuss the contribution of each collision term in what follows, and note that $\sigv_{\chi \bar{\chi} \to \sm }$ includes both $2 \to 2$ annihilations (such as $\chi \bar{\chi} \to gg$) and $3 \to 2$ coannihilations (such as $\chi \bar{\chi}\, g \to gg$).
Depending on the dominant process, the DM abundance may be produced in the early universe in a variety of ways, including freeze-in and freeze-out processes.
For the convenience of the reader, throughout this section we provide analytical approximations for various cross-sections and rates. 
We note that all figures presented in this work are obtained via full numerical computations of the relevant quantities.

\subsection{Axion thermalization: \texorpdfstring{$\Gamma_{a\,\sm\to  \sm}$}{}} 
\label{sec:ALPth}

To understand the cosmological dynamics of the model, it is important to establish when the $a$ particles are in equilibrium with the SM bath and when they are not. 
Since the DM $\chi$ couples to the SM bath particles only via its interactions with the axion $a$, thermal decoupling of $a$ from the SM bath necessarily implies the thermal decoupling of $\chi$ as well.

The question of thermal axion production resulting from the axion coupling to gluons has been discussed in detail in the literature~\cite{Masso:2002np,Salvio:2013iaa}, including the use of thermal field theory to account for many body plasma effects (\eg{} thermal masses due to screening). 
Ref.~\cite{DEramo:2021psx} extends the analysis to much lower temperatures to include temperatures below the QCD phase transition, where $\pi \pi \rightarrow \pi a$ is the dominant process. 
The latter is calculated in chiral perturbation theory which is valid up to temperatures of $T\sim 62\,\MeV$~\cite{DiLuzio:2022tbb}. 
This issue is addressed using the prescription presented in Ref.~\cite{DEramo:2021psx}, where interpolation is used to match between the rate $\Gamma_{\pi \pi \rightarrow a \pi}$ at $T < 62\,\MeV$ and the rate $\Gamma_{\rm UV} = \Gamma_{gg \rightarrow a g} + \Gamma_{q \bar{q} \rightarrow a g} + \Gamma_{q g \rightarrow a q}+  \Gamma_{\bar{q} g \rightarrow a \bar{q}}$ at $T > 2\,\GeV$. 
The rates in the interpolation span $10$~orders of magnitude, 
therefore the axion production rates at temperatures between $\sim 60\,\MeV-2\,\GeV$ should be taken with caution. 
However, as we will see below, the final DM abundance is sensitive to this only for axion masses close to the QCD  phase transition. 
Additionally, we consider the rates of the QED Primakoff processes $\Gamma_{e \gamma \rightarrow e a} + \Gamma_{\bar{e} \gamma \rightarrow \bar{e} a } + \Gamma_{e \bar{e} \rightarrow \gamma a}$ provided in Refs.~\cite{Bolz:2000fu,Cadamuro:2011fd}, which yield the dominant contributions at $T \ll 62\,\MeV$.

Importantly, all the analyses mentioned above are oriented towards a massless pseudo-scalar---motivated by the QCD-axion---and assume a relativistic axion where $m_a \ll T$. 
For $m_a>T$, the dominant rate of axion production comes from decays and inverse decays. 
The axion decay rate was estimated in Ref.~\cite{Aloni:2018vki} (see Fig.~S1 therein). 
Above $m_a\gtrsim 2\,\GeV$, the decay can be calculated pertubatively to two gluons, while for $3 m_{\pi} \lesssim m_a\lesssim 2\,\GeV$, the decays occurs to hadrons and photons and for $m_a \lesssim 3m_{\pi}$ 
{predominantly} to photons.
We denote the axion decay rate to the bath as $\Gamma_{a \to \rm \sm}$.
 
We add up all the  calculated rates in both the relativistic and non-relativistic regimes and use these for $\Gamma_{a\;{ \rm \sm} \to \rm \sm}$. 
Note that $\Gamma_{a\;{ \rm \sm} \to \rm \sm}$ includes in it $\Gamma_{a \to \rm \sm}$. 
The rate is therefore a function of the axion mass and the temperature.
However, for $T \sim m_a$, the previously calculated rates in the relativistic and non-relativistic regimes are not expected to be precise. 
In particular, there may be small corrections to the total rate, where the finite temperature rate calculated by Refs.~\cite{Masso:2002np,Salvio:2013iaa,DEramo:2021psx} in the relativistic regime may still dominate over the decay rate even as $T$ approaches $m_a$. 
This effect mostly occurs at temperatures right above the QCD phase transition, where $\alpha_s$ is large, and higher order effects become more important.
   
\subsection{Axion decay to dark matter: 
\texorpdfstring{$\Gamma_{a \to \chi \bar{\chi}}$}{}}

For $m_a> 2m_\chi$, the axion can decay directly into the DM. 
If allowed, this will be the dominant source of DM freeze-in.
The axion decay rate into DM is given by
\beq 
    \Gamma_{a \to \chi \bar{\chi}}
=   m_a \frac{c_\chi^2  m_\chi^2 }{8\pi f_a^2}\sqrt{1-\frac{4m_\chi^2}{m_a^2}}\,.
\eeq
%

\subsection{Axion annihlation to dark matter: \texorpdfstring{$\sigv_{\chi \bar{\chi} \to a a }$}{}}

At leading order the DM annihilation into axions can be calculated from the tree-level $u-$ and $t-$channel diagrams. 
A detailed calculation of the thermally averaged cross-section related to this process is given in Appendix~\ref{app:chichitoaa}.
For $m_a,T \ll m_\chi$---relevant for much of the freeze-in and freeze-out parameter space---the thermally averaged cross is well-approximated by 
\beq
    \label{eq:chichiaa}
    \sigv_{\chi \bar{\chi} \to a a }
    \simeq \frac{c_\chi^4 m_\chi^2}{64\pi f_a^4} 
    \frac{T}{m_\chi}\,.
\eeq
%

\subsection{Dark matter bath annihilation and production: \texorpdfstring{$\sigv_{ \chi \bar{\chi} \to {\rm \sm}  }^{\rm sub}$}{}}
\label{sec: DM annihilation and production}

Alternatively, the DM can be produced directly from the bath, or annihilate directly into the bath, by bypassing on-shell axions. 
To leading order, the process proceeds via an off-shell axion $a^*$.
Thus, the thermally averaged cross section can be calculated directly from the axion production rates already presented: 
\begin{align}
    & \sigv_{ \chi \bar{\chi} \to {\rm \sm}}= \nonumber \\ 
    &\qquad \frac{1}{(n_\chi^{\rm eq})^2}\int \dfrac{dm_{a^*}^2 }{\pi}
    \dfrac{m_{a^*} \Gamma_{a^* \to \chi \bar{\chi}}\Gamma_{a^*\,{\rm \sm}\to {\rm \sm}} n_{a^*}^{\rm eq }}{(m_{a^*}^2-m_a^2)^2+(m_a\Gamma_{a})^2}\,,
    \label{dmtosm}
\end{align}
where $\Gamma_{a}$ is the total axion decay rate. 
Within the integral, the rates ($\Gamma$'s) should be evaluated at the off-shell axion mass, $m_{a^*}$. 
A full derivation can be found in Appendix~\ref{app:dmviaa}.

In the narrow-width approximation, where the axion is produced on-shell, one can verify that 
\beq
    \sigv_{ \chi \bar{\chi} \to {\rm \sm} }^{\rm on-shell} \,\, 
    \Rightarrow \,\, \frac{n_a^{\rm eq}}{(n_\chi^{\rm eq})^2}\;\Gamma_{a\,  {\rm \sm}\to {\rm \sm}}\; {\rm BR}({a \to \chi \bar{\chi}}) \, ,
 \label{eq:onshellsmtochi}
\eeq
where $\Br({a \to \chi \bar{\chi}})$ is the branching ratio for the axion decay into DM.
The possibility that the DM produces an on-shell axion which then decays back to the SM bath is, in fact, already included in the other terms in the Boltzmann equations and must then be subtracted from this term to avoid double counting.
The on-shell subtracted thermally averaged cross section  is therefore given by
\beq
    \sigv_{ \chi \bar{\chi} \to \sm}^{\rm sub} 
=   \sigv_{ \chi \bar{\chi} \to \sm} 
    - \sigv_{\chi\bar{\chi}\to\sm }^{\rm on-shell} \, .
\eeq
Far from the axion resonance and for $T\ll m_\chi$, we have the simple relationship:
\beq
\left. \sigv_{\chi\bar{\chi}\to\sm}^{\rm sub} \right|_{T\ll m_\chi}
\!\!\!\!\simeq
\frac{16 c_\chi^2 m_\chi^6}{(m^2_a-4m_\chi^2)^2 f_a^2} \frac{\Gamma_{a^* \to \rm \sm}(2m_\chi)}{(2m_\chi)^3}\,.
\label{eq:chichitosmsigmav}
\eeq

In the other regime, for high temperatures, $T\gg m_a ,m_\chi$---which will be relevant for freeze-in of the DM---the thermally averaged cross section takes on the simple form
\begin{align}
    \label{eq:chichilargeT}
 &   \left.\sigv_{ \chi \bar{\chi} \to {\rm \sm} }^{\rm sub} \right|_{T\gg m_a > 2 m_\chi}\nonumber\\
 & \quad \quad \quad
=   \frac{n_a^{\rm eq}}{(n_\chi^{\rm eq})^2} \frac{2 c_\chi^2 m_\chi^2 T^3}{ \pi^2f_a^2 }
     \frac{\Gamma_{a^* {\rm \sm} \to {\rm \sm}}(\tilde m_{a} )}{ \tilde m_{a}^3 }\, ,
\end{align}
where ${\Gamma_{a^* {\sm} \to {\sm}}(\tilde{m}_{a} )}/{ \tilde{m}_{a}^3 }$ should be evaluated at $\tilde  m_{a} \simeq 1.8 T$ where the integral in Eq.~\eqref{dmtosm} is dominated. 
As this value falls within the region of uncertainty of the rate $\Gamma_{a^*\sm\to\sm}$, we approximate it by using only the decay rate contribution $\Gamma_{a^* \sm\to\sm} \simeq \Gamma_{a^* \to\sm}$. 
This introduces an $\mathcal{O}(10-100)$ uncertainty in Eq.~\eqref{eq:chichilargeT}, which corresponds to a correction of  $\mathcal{O}(1-3)$ in the required $f_a$ for UV-dominated freeze-in through this process.

\subsection{Meson decay: \texorpdfstring{$\Gamma_{P \to \chi \bar{\chi}}$}{}}

The neutral pseudoscalar mesons $P=\pi^0,\eta, \eta'$ can decay to DM via 
 mixing with the axion.
The mixing can be calculated in the chiral Lagrangian. 
Following Ref.~\cite{Aloni:2018vki} we find the simple relation
\bea
    \Gamma_{P\rightarrow \chi \bar{\chi}} 
    \simeq 
    \abs{\theta_{aP}^2} \Gamma_{a^* \to \chi \bar{\chi}}\,,
\eea
where $m_a^* = m_P$. 
In the numerical computations presented in this work, the mixing angle $\theta_{aP}$ is determined  by diagonalizing the kinetic and mass terms of $a$ and $P$. 
We note that our obtained mixing angle is inaccurate when the mass difference between $a$ and $P$ is of order their decay widths or smaller.

\section{Freeze-in}
\label{sec:FI}

Having addressed the relevant interactions of the DM, axion and SM bath, we now move to discuss how the DM abundance is set in the early universe, starting with freeze-in. 
Freeze-in is a dynamic process for producing a thermal relic of DM that assumes the DM abundance at early times to be negligibly small compared to its equilibrium abundance~\cite{Hall:2009bx}. 
A simple realization of such a case is post-inflationary reheating that reheats the SM bath, but not the DM.  
We take the initial temperature at which the axion and the DM can begin to be produced to be the reheating temperature $\TRH$, although we remain agnostic to the exact mechanism leading to such initial conditions. 
We consider that $\TRH < 8\pi f_a$, such that the production is not sensitive to the physics above the cut-off. 

As the $aG\tilde{G}$ coupling we consider is non-renormalizable, freeze-in of $a$ and $\chi$ is prone to being UV-dominated~\cite{Hall:2009bx},    largely determined by physics at $\TRH$. 
In general, when the majority of $\chi$ particles are produced directly from the SM bath, the production of $\chi$ will be dominated near $\TRH$. 
Likewise, if the axion never thermalizes, the production will depend on the UV-sensitive frozen in axion abundance. 
Otherwise, if the $\chi$ particles are produced from thermal axions or meson decays, then the production is determined by the renormalizable $a\chi\bar{\chi}$ interaction and production will be IR dominated, mostly occurring at $T=\max(m_a,m_\chi)$. 
In this section, we shall identify the processes governing UV-dominated and IR-dominated freeze-in, and describe the validity of each regime.
\begin{figure*}[th!]
	\centering
	\includegraphics[width=0.5\textwidth]{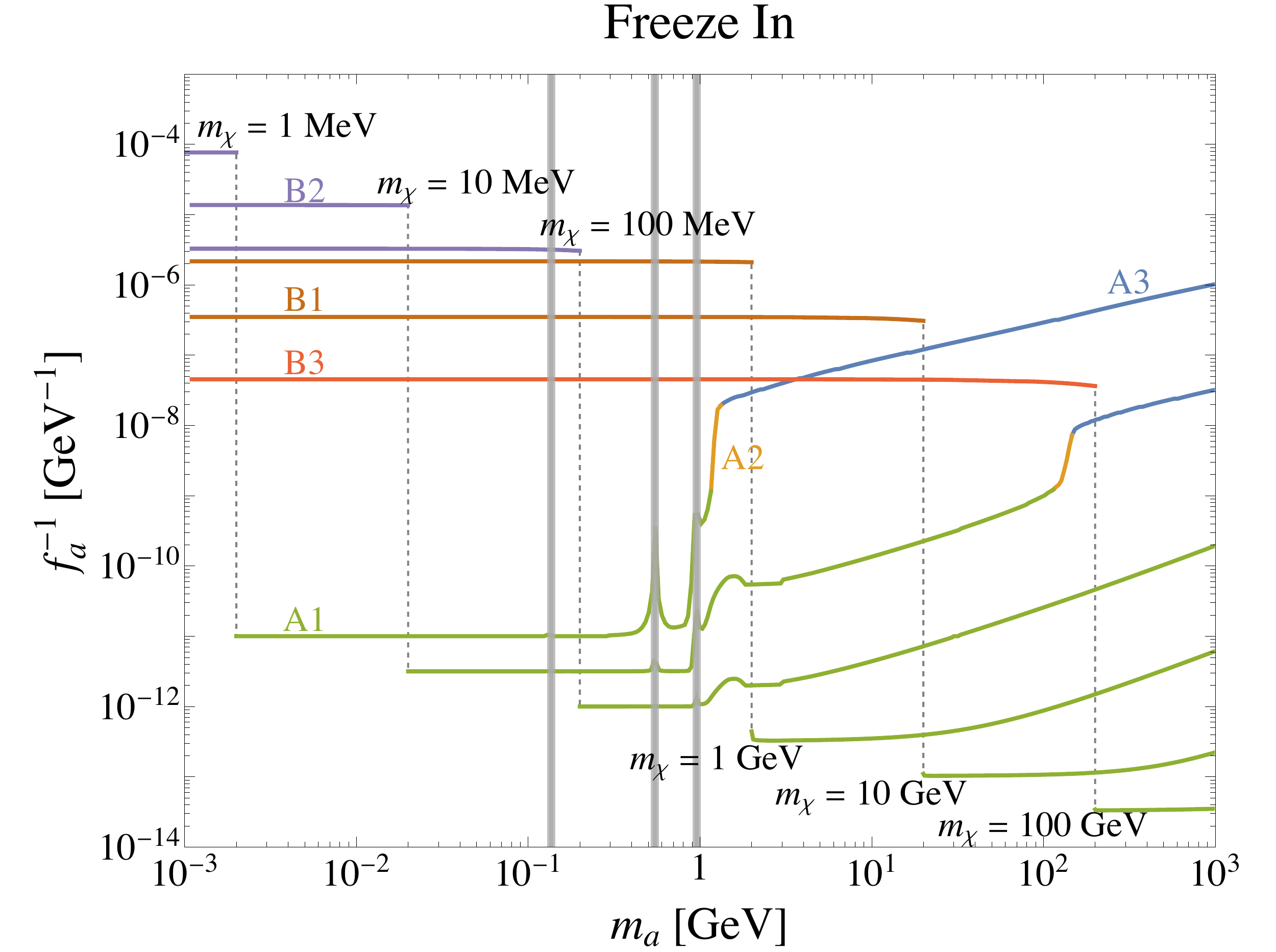}\hfill
 \includegraphics[width=0.5\textwidth]{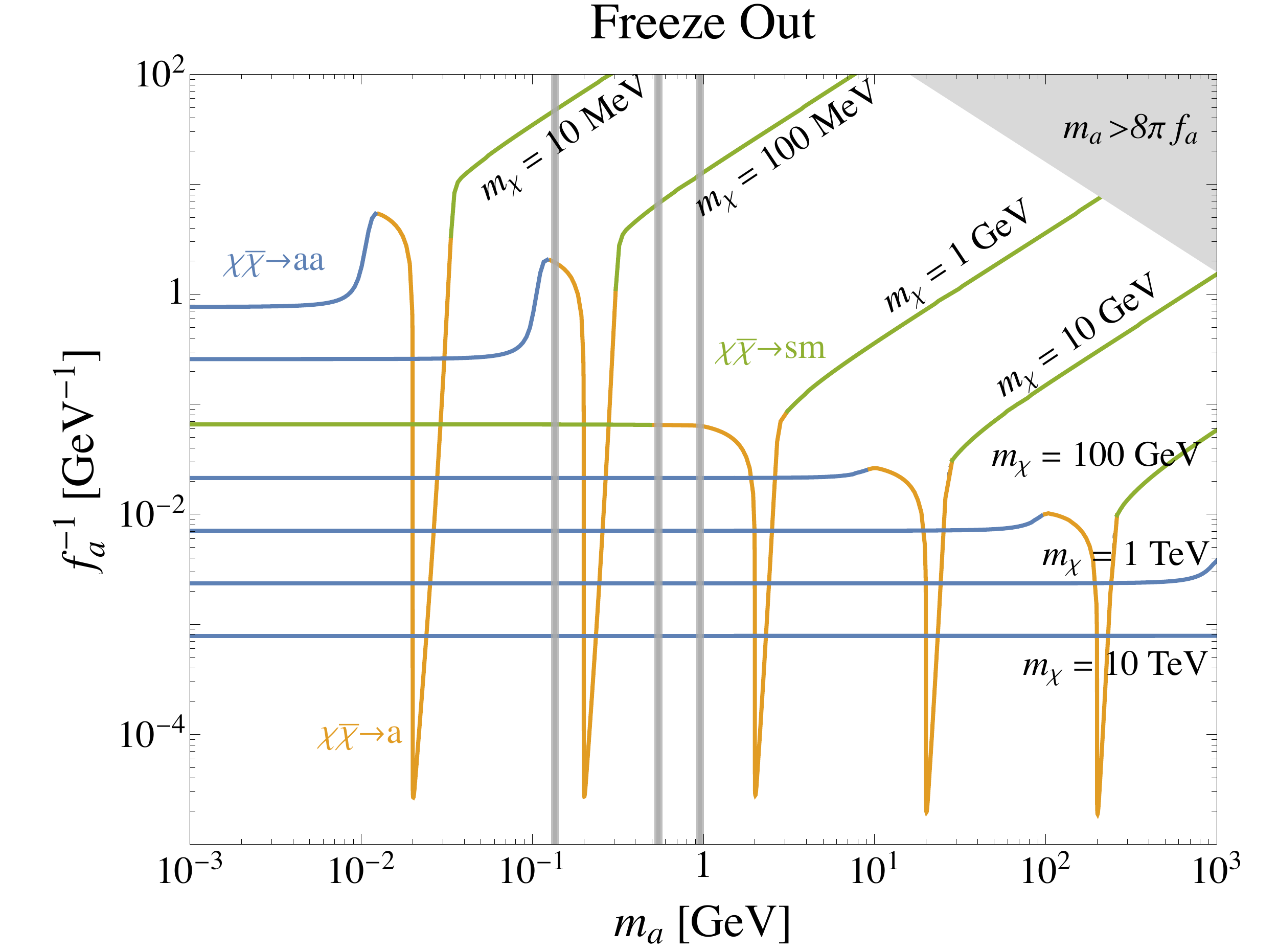}
	\caption{{\bf Dark matter freeze in and freeze out.}
 {\it Left:} 
 {Contours of fixed $m_{\chi}$ indicate the values of $\left( m_{a}, f_{a} \right)$ at which the observed DM relic abundance is obtained through freeze-in. The 
different colors illustrate the six different freeze-in regimes. For the invisibly decaying axion: regime {\tt A1}~(\textbf{{\color[rgb]{0.56, 0.69, 0.19}green}}), in which the axion never thermalizes with the bath; regime {\tt A2}~(\textbf{{\color[rgb]{0.88, 0.61, 0.14}orange}}), in which the axion thermalizes but decouples while relativistic; regime {\tt A3}~(\textbf{{\color[rgb]{0.36, 0.50, 0.71}blue}}), in which the axions decay in equilibrium with the bath. 
For the visibly decaying axion: regime {\tt B1}~(\textbf{{\color[rgb]{0.77, 0.43, 0.10}brown}}), UV-dominated bath production; regime {\tt B2}~(\textbf{{\color[rgb]{0.53, 0.47, 0.70}purple}}), pseudo scalar decay, and regime {\tt B3}~(\textbf{{\color[rgb]{0.92, 0.39, 0.21}red}}) axion annhilations.   
{\it Right:}  Contours of fixed $m_{\chi}$ indicate the values of $\left( m_{a}, f_{a} \right)$ at which the observed DM relic abundance is obtained through freeze-out. 
 The different colors along each solid curve illustrate each of the three freezeout regimes in our model. \textbf{{\color[rgb]{0.36, 0.50, 0.71}Blue}}: $\chi \bar{\chi} \rightarrow a a$ controls freezeout for $m_{\chi} \gtrsim m_{a}$ (except where meson resonances enhance annihilation to SM); \textbf{{\color[rgb]{0.88, 0.61, 0.14}orange}}: axion resonance $\chi \bar{\chi} \rightarrow a \rightarrow \text{SM}$ controls $m_{\chi} \lesssim m_{a} \lesssim 3 m_{\chi}$; \textbf{{\color[rgb]{0.56, 0.69, 0.19}green}}: $\chi \bar{\chi} \rightarrow \text{SM}$ controls $m_{a} \gtrsim 3 m_{\chi}$. 
  {\it In both panels}, grey vertical lines mask the regions where our numerical calculations do not accurately describe the effects of meson resonances due to finite resolution. 
} 
 }
 \label{fig:DM}
\end{figure*}

For the axion portal, freeze-in of the DM can be separated into two regimes. 
The first is when $m_a \ge 2 m_\chi$. 
In this case, freeze-in will always be dominated by the decay $a\to\chi \bar{\chi}$ regardless of the reheating temperature and whether or not the axion reaches chemical equilibrium with the bath. 
The second regime is when the decay is kinematically forbidden, $m_a < 2 m_\chi$. 
Here freeze-in will be dominated by axion-axion annihilation, SM bath annihilition or via meson decay into $\chi$ particles, depending on the reheat temperature and axion mass. 
Of the processes mentioned, only SM bath annihlation $\sm\to \chi \bar{\chi}$ is UV-dominated, whereas the rest are dominated at temperatures similar to the mass of the constituents.

The left panel of Fig.~\ref{fig:DM} presents solutions to the Boltzmann equations along contours of fixed $m_{\chi}$ which produce the observed DM relic abundance through freeze-in. 
Below we discuss the solutions to the Boltzmann equations and give analytical estimates of the results. 
For the figure we chose $\TRH=10\, \TeV$.

\subsection{\texorpdfstring{$m_a > 2 m_\chi$}{}: \texorpdfstring{$a\to \chi\bar{\chi}$}{} freeze-in}
We begin with the case in which the axion can decay into the DM. 
Here the freeze-in of the DM proceeds via the production of the axion and then its decay into DM. 
There are three relevant regimes, dependent on the thermal history of the axion:
\begin{enumerate}
    \item {\it The axion never thermalizes with the bath.}
    If the axion is too feebly interacting to thermalize with the bath, then its abundance will be populated by freeze-in. 
    This process is UV-dominated so it reaches an asymptotic co-moving number density near $\TRH$, given simply by integrating over the rate of production,
    \bea 
        \label{eq:axion freeze in}
        Y_a^{\rm FI}
    &=&   \int_0^{\TRH} dT \frac{n_a^{\rm eq}(T)\Gamma_{a\,\sm\to\sm}(T)}
        {T H(T) s(T)}\no\\
        &\simeq& 
        \frac{135 \sqrt{5} \mPl \Gamma_{a\,\sm \to\sm}(\TRH) }
        {\sqrt{2} \pi^5 \sqrt{g_\star(\TRH)} g_{\star s}(\TRH)\TRH^2}\,,
    \eea
    where $g_\star$\,($g_{\star s}$) is the effective number of relativistic\,(entropy) degrees of freedom.  
    The integral was performed assuming $\Gamma_{a\,\sm\to\sm} \propto T^3$. 
    The subsequent $\chi$ abundance is just the fraction of these axions that decay into DM,
    \begin{equation}
        Y_{\chi}(\infty) = 2 Y_a^{\rm FI} \Br(a \to \chi \bar{\chi})\, .
    \end{equation}
This regime is demonstrated in the left panel of Fig.~\ref{fig:DM} by the green parts of the curves, labeled {\tt A1}. 
The shape is controlled by the branching ratio of axions to the DM, and the shape can be matched to Fig.~\ref{fig: branching ratio to inv} in Appendix~\ref{app: inv BR}.
\item {\it The axion thermalizes but decouples while relativistic.} 
Decoupling from the bath occurs when 
\beq
    \label{axiontherm}
    \Gamma_{a\,\sm\to\sm}
    \simeq  
    H \Big|_{T=T_{\rm dec}}\,.
\eeq
After $T_{\rm dec}$, the comoving abundance is fixed to its equilibrium value at $T_{\rm dec}$ until it decays. The $\chi$ abundance is then given by 
\beq
    \label{FIrelaxion}
    Y_\chi(\infty) 
=   2 Y_a^{\rm eq}(T_{\rm dec}) \Br(a \to \chi \bar{\chi})\,.
\eeq 
Corrections to the axion production rate, $\Gamma_{a\,\sm\to\sm}$, in the region interpolated between the chiral Lagrangian calculation and the QCD calculation may change $T_{\rm dec}$ in Eq.~\eqref{axiontherm}.  
In terms of the relic abundance of the DM, the effect will be a change in $g_{\star s}$ at the time of axion decoupling that appears in $Y_a^{\rm eq}(T_{\rm dec})$. 
Near the QCD phase transition, this can alter the relic abundance up to a factor of six, which
would correspond to $\mathcal{O}(1)$ corrections to the $m_\chi$, $m_a$ and $f_a$ values needed to match the observed abundance. 
This regime is shown in the left panel of Fig.~\ref{fig:DM} by the orange parts of the curves labeled {\tt A2}.
\item {\it The axions  decay in equilibrium with the bath.} This occurs when the axions stay thermalized until they become non-relativistic, and then decay.
The relic abundance of DM is then given simply by the rate of thermal axion decays into the DM,
\bea
    Y_\chi(\infty)
    &=& 
    \int_0^{\TRH} dT  \frac{n_a^{\rm eq}(T)\Gamma_{a\to \chi \bar{\chi} }}
    {T H(T) s(T)}\no\\
    &\simeq& 
    \frac{135\sqrt{5} c_\chi^2 \mPl  m_\chi^2}
    {2\sqrt{2} \pi^6 \sqrt{g_{\star}(m_a)} g_{\star s}(m_a) m_a f_a^2} \, .
\eea
This regime is shown in the left panel of Fig.~\ref{fig:DM} by the blue curves labeled {\tt A3}.  
\end{enumerate}
%

\subsection{\texorpdfstring{$m_a \le 2 m_\chi$}{}: axion annihilation freeze-in}

Next, we consider the case where the axion decay to DM is forbidden. 
In this case, there are three different sources of DM freeze-in production: 
\begin{enumerate}
\item {\it Production directly from the bath.}    
For much of the parameter space considered here, the dominant contribution of DM is direct production from the bath. This is a UV-dominated process and will be sensitive to the reheat temperature. 
The freeze-in abundance is then 
\bea
    Y_\chi(\infty)
&=& \int_0^{\TRH} dT \frac{{(n_\chi^{\rm eq})}^2
    \sigv_{\chi \bar{\chi} \to  \sm }}{T H(T) s(T)}\no \\
    &\simeq& 
    \left. T\frac{{(n_\chi^{\rm eq})}^2\sigv_{\chi \bar{\chi} \to \sm }}
    {T H(T) s(T)} \right|_{T=\TRH}  \,.
    \eea
This regime is shown in the left panel of Fig.~\ref{fig:DM} by the brown curves labeled {\tt B1}. 
This will be the dominant source of production of the DM for masses  between the curves $m_\chi=m_{\eta'}/2 = 478\, \MeV$ and $m_\chi \simeq 20\, \GeV$. 
The $m_\chi \simeq 20\, \GeV$ boundary is a result of the choice of $\TRH =10\,\TeV$. 
Increasing $\TRH$ would enlarge this regime. 
\item {\it Meson decays.} 
The decay of the psuedo-scalar mesons $\pi^0$, $\eta$ and $\eta'$ freezes-in the majority of DM when the decay is kinematically allowed. 
For this case,
\begin{align}
	Y_{\chi}(\infty) 
= & \int_0^{T_{\rm QCD}} dT \frac{n_P^{\rm eq}(T) 
    \sum_{P = \pi^0,\eta,\eta'}\Gamma_{P\to\chi\bar{\chi}}}
    {T H(T) s(T)}\no\\
    \simeq & 
    \sum_{P = \pi^0, \eta, \eta'}\frac{135 \sqrt{5} }{8 \pi^{11/2}} \frac{\mPl \Gamma_{P \to \chi \bar{\chi}}}{\sqrt{g_{\star}(m_{\pi})} g_{\star s}(m_{\pi}) m_P^2 }\no\\
	&\qquad \times  \pqty{\frac{m_P}{\Lambda_{\rm QCD}}}^3 
    K_{3} \pqty{ \frac{m_P}{\Lambda_{\rm QCD}}}\,,
\end{align}
where $T=\LamQCD$ is the QCD phase transition temperature and $K_3(x)$ is the third modified Bessel function of the second kind. 
To achieve this nice closed form we have evaluated the effective number of degrees of freedom of the SM bath at $m_{\pi}$, since freeze-in through such decays is dominated by temperatures just below $\LamQCD \sim m_\pi$. 
 This production is largely insensitive to the reheat temperature, unlike the direct bath production described above that grows with reheat temperature.
This shown in the left panel of Fig.~\ref{fig:DM} by the purple curves labeled~{\tt B2}.
\item {\it Axion annihilations.} 
At low enough reheat temperatures and for DM masses large enough ($m_{\chi} \gtrsim 20\, \GeV$ for $\TRH= 10\,\TeV$ as shown in the left panel of Fig.~\ref{fig:DM}), the majority of the DM freezes in from axion annihilation. 
This contribution will only compete with the direct production of DM from the SM bath when the axions are thermalized. 
Therefore, we assume a thermal distribution of axions when calculating the abundance from this process. 
Freeze-in here is IR-dominated and most of the DM is produced soon after the axions become non-relativistic and deplete.  
Therefore, we can integrate Eq.~\eqref{eq:chichiaa} to approximately obtain the freeze-in abundance:
\bea
    Y_\chi(\infty)
&=& \int_0^{\TRH} dT  \frac{{(n_\chi^{\rm eq})}^2
    \sigv_{\chi \bar{\chi} \to a a }}{T H(T) s(T)}\no \\
    &\simeq& 
    6\times 10^{-4} \frac{c_\chi^4 m_{\rm pl } m_\chi^3}{f_a^4 \sqrt{g_{\star}(m_\chi)} g_{\star s}(m_\chi)}\,,
\eea
where we numerically perform the integral for $m_a < 2 m_\chi$ and $\TRH \gg m_a,m_\chi$. 
This is outlined by the red curves in the left panel of Fig.~\ref{fig:DM} labeled {\tt B3}.
\end{enumerate}

\section{Freeze-out}
\label{sec:FO}

In this section we consider freeze-out, where the DM is in thermal equilibrium with the SM bath at early times and its relic abundance is set by the decoupling of DM number-changing processes. 
In all regions of parameter space where the $\chi$ freezes out from the bath, the axions are also in equilibrium with the bath. 
Therefore, one only needs to study the 
Boltzmann equation of
\begin{align}
    \label{eq: FO boltzmann eq}
	\dot{n}_\chi \!+\! 3 H n_\chi 
=&  -\pqty{
    \sigv_{\chi \bar{\chi} \rightarrow aa} 
    \!+\! \sigv^{\rm sub}_{\chi \bar{\chi} \rightarrow {\rm \sm}} 
    \!+\! \sigv_{\chi \bar{\chi}\rightarrow a}
    } \nonumber\\
    & \times \pqty{ n_\chi^2-n_\chi^{\rm eq\,2} }\,.
\end{align}
The parameter space where the DM relic abundance is obtained via freeze-out can be understood in three main regimes, each corresponding to the dominance of a different term in Eq.~\eqref{eq: FO boltzmann eq}.  
The right  panel of Fig.~\ref{fig:DM} presents solutions to the Boltzmann Eq.~\eqref{eq: FO boltzmann eq} for  fixed $m_\chi$ leading to the observed DM relic abundance today, $m_\chi Y_\chi \simeq 0.43\,\eV$ through freeze out.
The different colored segments along each curve illustrate each of the three freeze-out regimes:
\begin{enumerate}
\item {\it DM annihilation to axions}. $\chi\bar{\chi}\to aa$ dominates for $m_\chi > m_a$ (with the exception of $2 m_\chi \sim m_{\pi^0}, m_\eta, m_{\eta'}$ where mesons resonances enhance annihilation to SM); depicted in blue.
For $m_{\chi} < m_{a}$, the annihilation into axions becomes kinematically suppressed.  
\item {\it Axion resonance (inverse decay)}. $\chi \bar{\chi}\to a \to\sm$ dominates for $2m_\chi \lesssim m_a \lesssim 3m_\chi$; depicted in orange. 
\item {\it Annihilation into bath particles}. $\chi\bar{\chi}\to \sm$ through an off-shell axion dominates for $m_a \gtrsim 3m_\chi$; illustrated in green.  When $2 m_\chi$ is close to hadronic resonances, the annihilation into bath particles can dominate for $m_a < 3m_\chi$ as well.
\end{enumerate}

Following Ref.~\cite{Frumkin:2022ror}, the mass-coupling relationship to match the observed abundance can be obtained semi-analytically. 
For $m_\chi>m_a$, using Eq.~\eqref{eq:chichiaa} we find
\beq 
    \label{eq:ronny}
    m_\chi \simeq 11\,\TeV \times \pqty{\frac{f_a}{\TeV} }^2 \, .
\eeq
For $m_\chi\ll m_a$  (far from the axion resonance), by using Eq.~\eqref{eq:chichitosmsigmav} we find
\beq
    m_\chi 
    \simeq 
    2\,\TeV \left(\frac{f_a m_a}{\TeV^2}\right)^{\frac{2}{3}} 
    \left(\frac{f_a^2\Gamma_{a\to \rm \sm}/m_a^3}{10^{-5}}\right)^{-\frac{1}{6}}
\eeq
where we have taken $g_{\star}(m_\chi)=g_{\star s}(m_\chi)=106.75$.

For $m_a \simeq 2m_\chi$, the annihilation is dominated by the axion resonance. 
Using Eq.~\eqref{eq:onshellsmtochi}, we find the relic abundance is mostly $m_\chi$ and $m_a$ independent in this case (up to $g_\star$ and $g_{\star s}$ corrections) and 
\beq
    \label{eq:resonance freeze out coupling}
    f_a\simeq 30\,\TeV\,,
\eeq
reproduces the observed abundance.

\begin{figure*}[th!]
	\label{fig: experimental constraints}
	\centering
	\includegraphics[width=0.495\textwidth]{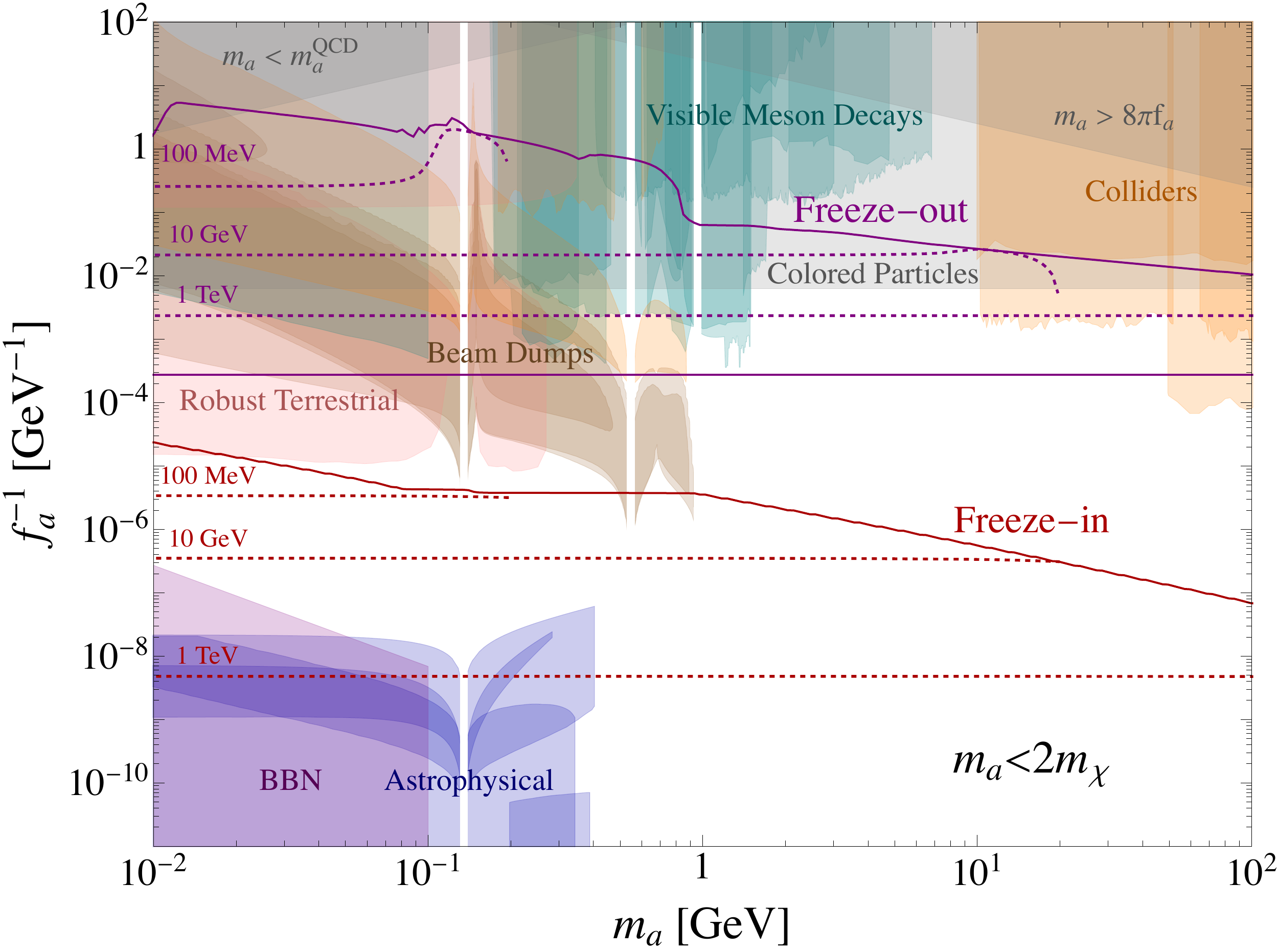}
	\hfill\includegraphics[width=0.495\textwidth]{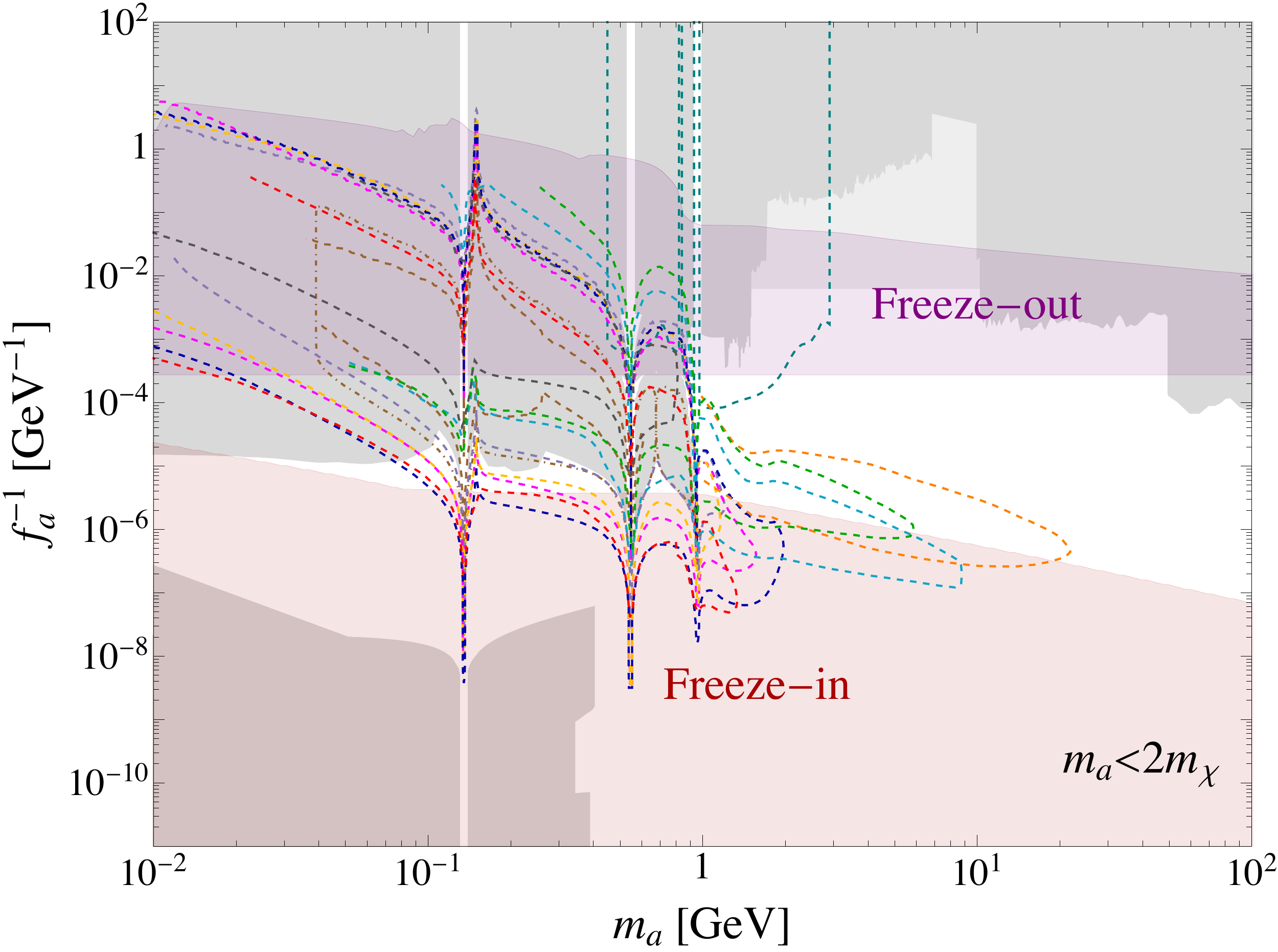}
	\caption{{\bf Axion coupling to gluons for $\boldsymbol{m_a < 2m_\chi}$.}
	{\it Left:} Constraints on the coupling: robust terrestrial bounds~\cite{NA62:2021zjw,Bauer:2021mvw,Goudzovski:2022vbt}~(\textbf{{\color[rgb]{1,.5,.5}pink}}), 
    beam dumps \cite{Jerhot:2022chi, Blumlein:1990ay,Blumlein:2013cua,CHARM:1980zcj,CHARM:1983vkv,CHARM:1985anb,Bjorken:1988as,Riordan:1987aw,Afik:2023mhj}~(\textbf{{\color[rgb]{0.6, 0.4, 0.2}brown}}), 
    meson decays \cite{Bauer:2021mvw,LHCb:2016awg,NA62:2014ybm,E949:2005qiy,BaBar:2011kau,Chakraborty:2021wda,Belle:2013nby, BaBar:2008rth, BaBar:2011vod, BESIII:2022rzz}
    ~(\textbf{{\color[rgb]{0,.5,.5}turquoise}}), 
    colliders~\cite{CMS:2014mvm, CMS:2015ocq, CMS:2017yta, CMS:2017dcz, CMS:2019nrx, CMS-PAS-HIG-14-037, CMS-PAS-HIG-17-013, ATLAS:2012fgo,ATLAS:2014jdv, ATLAS:2017cvh,ATLAS:2022abz,AxionLimits,Mimasu:2014nea, GlueX:2021myx,Mariotti:2017vtv,CMS:2021juv,Mitridate:2023tbj}~(\textbf{{\color[rgb]{1, 0.5, 0}orange}}), 
    BBN  \cite{Kawasaki:2017bqm, Kawasaki:2020qxm}~(\textbf{{\color[rgb]{0.5, 0, 0.5}purple}}), 
    astrophysical \cite{Lella:2022uwi}~(\textbf{{\color[rgb]{0, 0, 0.66}dark blue}}) and new colored particles \cite{ATLAS:2019fgd,CMS:2018mts,ATLAS:2017jnp,CMS:2018mgb,ATLAS:2021mdj}~(\textbf{{\color[rgb]{0.5,0.5,0.5}gray}}).  
    EFT constraints are indicated by \textbf{{\color[rgb]{0.5,0.5,0.5}gray}} shaded regions.
    The dashed curves are numerical solutions to the Boltzmann equations giving the correct DM abundance today for the various labeled DM masses. 
    {\it Right:} The freeze-in~(\textbf{{\color[rgb]{0.66, 0, 0}dark red}}) and freeze-out~(\textbf{{\color[rgb]{0.5, 0, 0.5}purple}}) phases in the theory considered here as one varies the DM mass, with regions excluded by existing constraints~(\textbf{{\color[rgb]{0.5,0.5,0.5}gray}}) and regions excluded only by the colored particles bound~(\textbf{{\color[rgb]{0.67,0.67,0.67}light gray}}).
    The upper limit of both regions stems from $2 m_\chi \gtrsim m_a$; for the freeze-out region the lower limit arises from $m_\chi = 8 \pi f_a$ which is comparable to the unitarity bound. Dashed curves show the projected sensitivities of future experiments: 
    DUNE-ND~(\textbf{{\color[rgb]{1, 0, 0} red}})~\cite{Kelly:2020dda}, 
    Belle-II~(\textbf{{\color[rgb]{0, 0.5, 0.5}turquoise}})~\cite{Chakraborty:2021wda}, 
    LHC track trigger~(\textbf{{\color[rgb]{1, 0.5, 0}orange}})~\cite{Hook:2019qoh}, 
    MATHUSLA~(\textbf{{\color[rgb]{0.068, 0.64, 0.78}light blue}})~\cite{Chou:2016lxi, Kelly:2020dda}, 
    SHiP~(\textbf{{\color[rgb]{0.0, 0.0, 0.67}dark blue}}) \cite{SHiP:2015vad, Jerhot:2022chi},
    NA62-LS3~(\textbf{{\color[rgb]{1,0,1}magenta}})~\cite{NA62:2017rwk, Jerhot:2022chi},
    SHADOWS~(\textbf{{\color[rgb]{1,0.75,0}yellow}})~\cite{Baldini:2021hfw, Jerhot:2022chi},
    DarkQuest~(\textbf{{\color[rgb]{0.52,0.47,.7}light purple}}) ~\cite{Batell:2020vqn, Jerhot:2022chi},
    CODEX-b~(\textbf{{\color[rgb]{0,0.67,0}green}}) ~\cite{Gligorov:2017nwh, Kelly:2020dda}, 
    FASER~(\textbf{{\color[rgb]{.33,.33,.33}gray}})~\cite{Feng:2018pew, FASER:2018eoc, Kelly:2020dda} and KOTO~(\textbf{{\color[rgb]{.6,.4,.2}brown}}), along with KOTO2~(\textbf{{\color[rgb]{.6,.4,.2}brown dot-dashed}})~\cite{Afik:2023mhj,Yamanaka:2012yma,Aoki:2021cqa}.
    }
    \label{fig:exp}
\end{figure*}

\begin{figure*}[th!]
	\centering
	\includegraphics[width=0.49\textwidth]{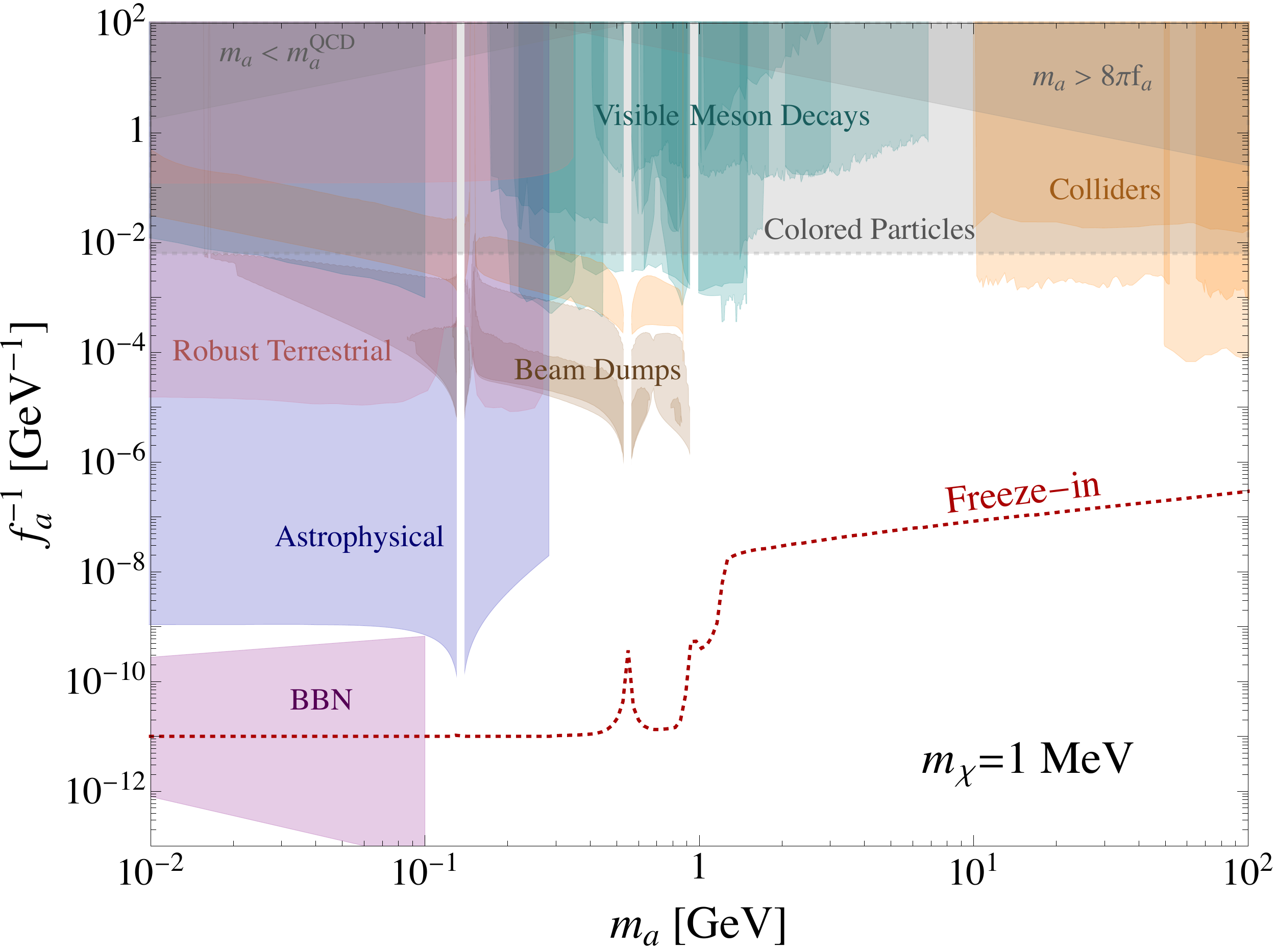}
        \includegraphics[width=0.49\textwidth]{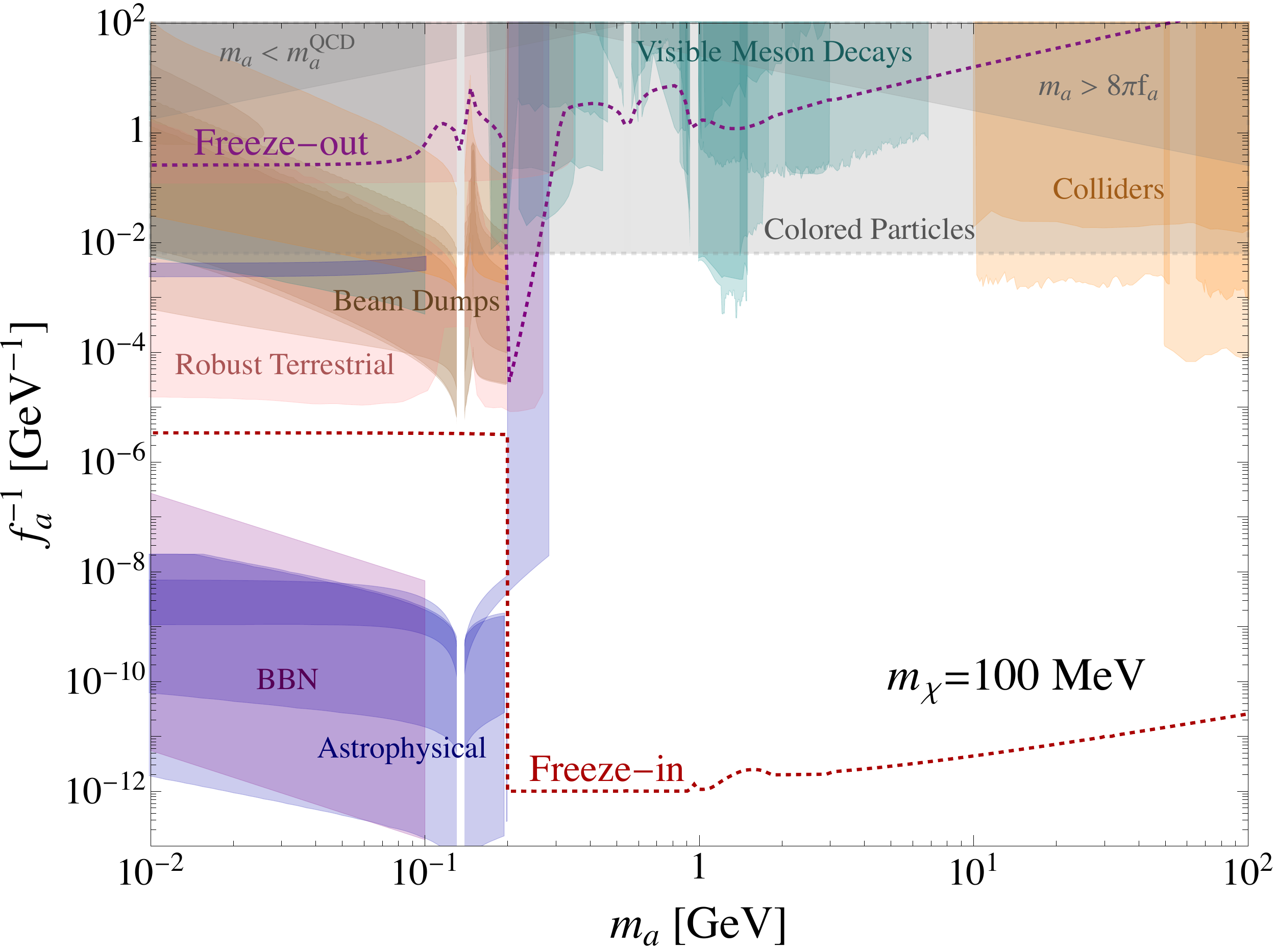}
        \\
	\includegraphics[width=0.49\textwidth]{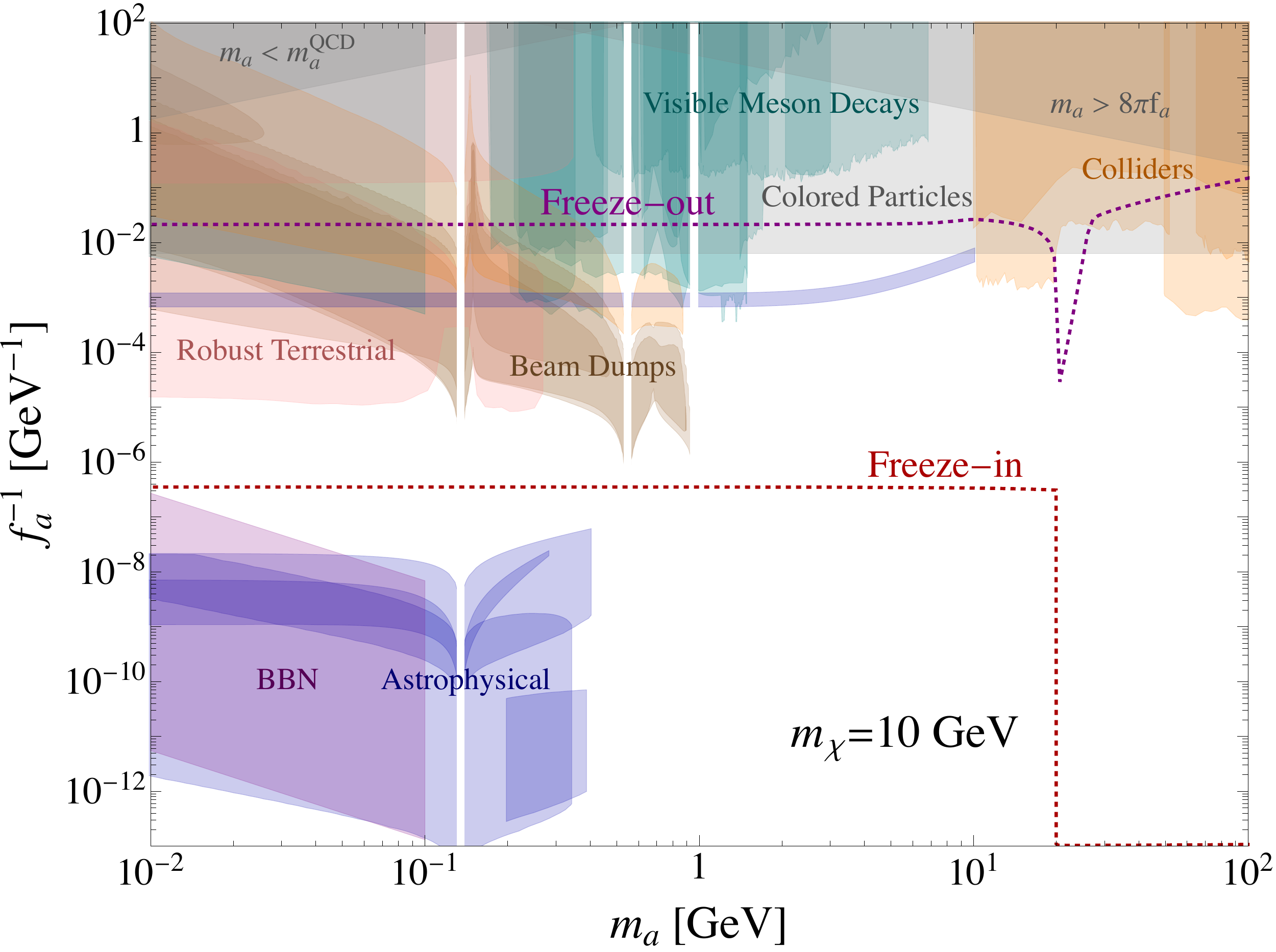}
        \includegraphics[width=0.49\textwidth]{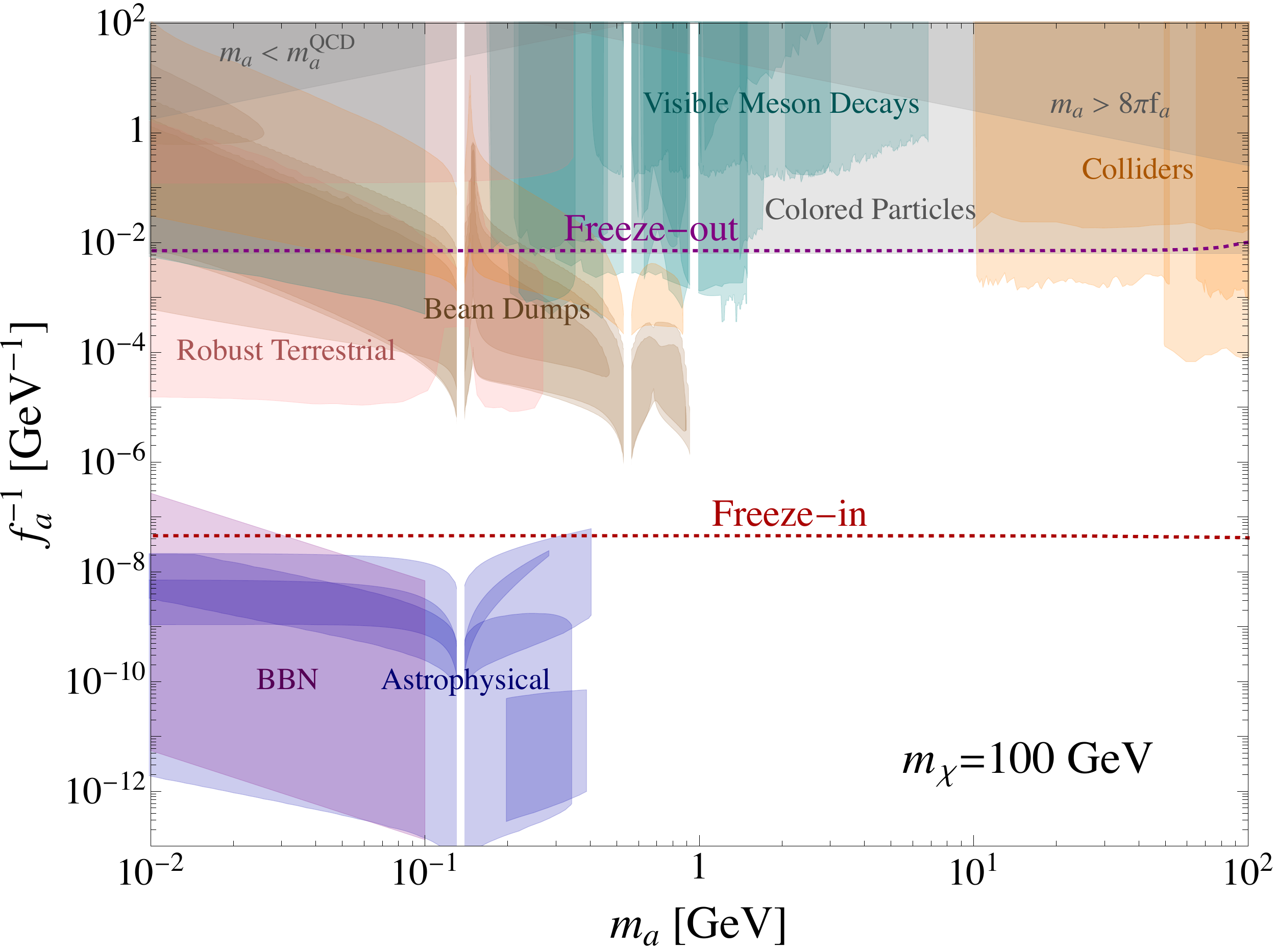}
	\caption{{\bf Axion coupling to gluons at fixed $m_\chi$, including $\boldsymbol{m_a>2 m_\chi}$.} We show 
 constraints on the coupling for various DM 
 masses: MeV~({\it upper left}), 100\,MeV~({\it upper right}), 10 GeV~({\it lower left}), 100 GeV~({\it lower right}). 
 The colors of the shaded regions indicate the different types of constraints: 
 robust terrestrial bounds~\cite{NA62:2021zjw,Bauer:2021mvw,Goudzovski:2022vbt}~
(\textbf{{\color[rgb]{1,.5,.5}pink}}), 
    beam dumps \cite{Jerhot:2022chi, Blumlein:1990ay,Blumlein:2013cua,CHARM:1980zcj,CHARM:1983vkv,CHARM:1985anb,Bjorken:1988as,Riordan:1987aw,Afik:2023mhj}~(\textbf{{\color[rgb]{0.6, 0.4, 0.2}brown}}), 
    meson decays \cite{Bauer:2021mvw,LHCb:2016awg,NA62:2014ybm,E949:2005qiy,BaBar:2011kau,Chakraborty:2021wda,Belle:2013nby, BaBar:2008rth, BaBar:2011vod, BESIII:2022rzz}
    ~(\textbf{{\color[rgb]{0,.5,.5}turquoise}}), 
    colliders~\cite{CMS:2014mvm, CMS:2015ocq, CMS:2017yta, CMS:2017dcz, CMS:2019nrx, CMS-PAS-HIG-14-037, CMS-PAS-HIG-17-013, ATLAS:2012fgo,ATLAS:2014jdv, ATLAS:2017cvh,ATLAS:2022abz,AxionLimits,Mimasu:2014nea, GlueX:2021myx,Mariotti:2017vtv,CMS:2021juv,Mitridate:2023tbj}~(\textbf{{\color[rgb]{1, 0.5, 0}orange}}), 
    BBN  \cite{Kawasaki:2017bqm, Kawasaki:2020qxm}~(\textbf{{\color[rgb]{0.5, 0, 0.5}purple}}), astrophysical \cite{Lella:2022uwi}~(\textbf{{\color[rgb]{0, 0, 0.66}dark blue}}) and new colored particles \cite{ATLAS:2019fgd,CMS:2018mts,ATLAS:2017jnp,CMS:2018mgb,ATLAS:2021mdj}~(\textbf{{\color[rgb]{0.67,0.67,0.67}light gray}}).  
    EFT constraints are indicated by \textbf{{\color[rgb]{0.5,0.5,0.5}gray}} shaded regions. 
The dashed curves indicate the numerical solutions to the Boltzmann equations for freeze out~(\textbf{{\color[rgb]{0.5, 0, 0.5}purple}})  and freeze in~(\textbf{{\color[rgb]{0.66, 0, 0}dark red}}) giving the correct DM abundance today.
	\label{fig:expinv}
	}
\end{figure*}

\section{Experimental constraints}
\label{sec:exp}

Having analyzed the DM relic abundance in various regions of parameter space, we move to address experimental constraints of the ALP-gluon portal discussed in this work.
The shaded regions of Figs.~\ref{fig:exp} and~\ref{fig:expinv} consolidate existing bounds on the axion coupling, $f_a^{-1}$, as a function of the axion mass, $m_a$.

The constraints fall into three categories: 
\begin{enumerate}[(i)]
\item \textit{Robust terrestrial bounds}: terrestrial experiments that are either based on invisible signatures or independent of the ALP decay final states~(shown in pink).
\item \textit{Visible terrestrial bounds}: terrestrial experiments that place constraints on visibly decaying ALPs~(shown in brown, turquoise and orange).
\item \textit{Cosmological and astrophysical bounds}: cosmological~(shown in purple) and astrophysical~(shown in dark blue) bounds.
\end{enumerate}
Details regarding the casting of the bounds, including in case of the invisibly decaying ALP, are provided in Appendix~\ref{app:casting constraints}.
Below we describe the various constraints.

In category~(i) of robust terrestrial bounds, measurement of the rare decay $K^+ \to \pi^+ \nu \bar{\nu}$ performed by the NA62 Collaboration~\cite{NA62:2021zjw} and analyzed in Ref.~\cite{Bauer:2021mvw} place constraints on ALP masses in the regime $m_a < m_{K^+} - m_{\pi^+}$. 
These constraints are constructed by bounding the number of ALPs that escape detection; as such they provide a robust limit independent of the final state, with the limit strengthening further when the ALP can decay invisibly. 
For larger couplings, the $K^+ \to \pi^+ a$ decay modifies the $K^+$ lifetime beyond the bound $\Br(K^+ \rightarrow \rm{sm + new\ physics}) \leq 3 \times 10^{-3}$ placed by Ref.~\cite{Goudzovski:2022vbt}.
Present data from $K_L \to \pi^0 \nu\bar{\nu}$ searches~\cite{KOTO:2020prk} places weaker bounds, and are not presented here.
(Here we do not show limits in the region $m_a < m_B - m_K$ derived in Ref.~\cite{Chakraborty:2021wda} from the analysis of the inclusive branching ratio of $\Br(b \to s a)$, nor the limits derived in Ref.~\cite{Bauer:2021mvw} for $m_a < m_t$ from the measurements of the chromomagnetic dipole moment of the top quark, as both should arise from an RG flow from a $f_a$-dependent UV scale.) 
 
The visible terrestrial bounds, category~(ii), contains constraints from beam dumps, meson decays and other collider searches. 
Ref.~\cite{Jerhot:2022chi} summarizes existing constraints from proton beam dumps, including limits from the NuCal~\cite{Blumlein:1990ay,Blumlein:2013cua} and CHARM~\cite{CHARM:1980zcj,CHARM:1983vkv,CHARM:1985anb} collaborations, which place some of the strongest limits in the region of  $m_a \lesssim 1\,\GeV$. 
We also consider constraints from the electron beam dumps E137 \cite{Bjorken:1988as} and E141 \cite{Riordan:1987aw}, presented in Ref.~\cite{Jerhot:2022chi}.
The beam dump bounds are complemented at larger couplings by the constraints derived from the CMS search for long-lived particles decaying in the muon endcap detectors \cite{CMS:2021juv}, as analyzed in Ref.~\cite{Mitridate:2023tbj}.
Additional bounds can be derived from $K$, $B$, $J/\psi$ and $\Upsilon$ meson decays. 
Here we have included the meson decays $B^+ \to K^+ a(\mu^+\mu_-)$, $K^+ \rightarrow \pi^+ a(\gamma\gamma)$ and $\Upsilon\rightarrow \gamma a$, summarized in Ref.~\cite{Bauer:2021mvw} based on measurements by the LHCb~\cite{LHCb:2016awg}, NA62~\cite{NA62:2014ybm}, E949~\cite{E949:2005qiy} and BaBar~\cite{BaBar:2011kau} collaborations and $B \to K a(3\pi)$, $B \to K a(\eta \pi \pi)$, $B \to K a(K K \pi)$ and $B \to K a(\phi\phi)$, summarized in Ref.~\cite{Chakraborty:2021wda} based on measurements performed by the Belle~\cite{Belle:2013nby} and BaBar~\cite{BaBar:2008rth,BaBar:2011vod} collaborations. 
We have further included a recent analysis by BESIII~\cite{BESIII:2022rzz} that places constraints on the $J/\psi \to \gamma a(\gamma \gamma)$ decay process. 
We learn that meson decays place the most stringent constraints on large ALP-SM couplings in the $m_a\sim 100\,\MeV- 7\,\GeV$ region.

For $m_a \gtrsim 10\,\GeV$, searches for di-jet~\cite{CMS:2017dcz} and di-photon~\cite{ATLAS:2012fgo, ATLAS:2014jdv, ATLAS:2017cvh, ATLAS:2022abz, CMS:2014mvm, CMS:2015ocq, CMS:2017yta, CMS:2019nrx,CMS-PAS-HIG-14-037, CMS-PAS-HIG-17-013} resonances at the LHC place the strongest limits on the parameter space. 
Here we have recast the data from Ref.~\cite{ATLAS:2022abz}, which is publicly available in Ref.~\cite{AxionLimits}, and the bounds analyzed in Ref.~\cite{Mariotti:2017vtv}, to encompass our ALP-gluon model. 
We do not show mono-jet and di-jet signatures of lower masses~\cite{Mimasu:2014nea} as they are weaker than the bounds we present. 
We have also cast bounds from GlueX which constrain the mass region $m_a \sim \mathcal{O}(100)\,\MeV$ \cite{GlueX:2021myx}.

As the analyses in category~(ii) all rely on visible final states, they are affected by the branching ratio $\Br(a \to {\rm visible})$ and the axion lifetime.
For $m_a > 2m_\chi$, the axion invisible branching ratio becomes dominant, weakening the sensitivity of these searches by a factor of $\sim \sqrt{1 -\Br(a \to \chi \bar{\chi})}$. 
Values of $\Br(a \to \chi \bar{\chi})$ for some representative DM masses are shown in Fig.~\ref{fig: branching ratio to inv} in Appendix \ref{app: inv BR}.
 
The constraints in catergory~(iii) come from cosmology and astrophysical probes, and include bounds from big bang nucleosynthesis~(BBN), 
supernovae~(SN), 
and proton-neutron star heating.
Ref.~\cite{Kawasaki:2020qxm} provides bounds on the synthesis of light elements during BBN in the presence of ALPs decaying to photons, for $m_a=10\,\MeV$ and  $100\,\MeV$.
 
SN bounds are taken from  Ref.~\cite{Lella:2022uwi} which considered implications of ALP nucleon, ALP-pion and ALP-photon interactions on various SN observables. 
We expect the SN bounds presented here to be complemented at stronger couplings by future dedicated analyses considering the effects of an ALP-nucleon coupling in addition to an ALP-photon coupling in low energy SN~\cite{Caputo:2022mah}. 
The SN bounds shown in this work can be improved by a dedicated numerical analysis and should be taken as an estimate of the region of exclusion. 
Note that we do not show the SN1987A bounds presented in \cite{Chang:2018rso, Ertas:2020xcc} as those works did not consider the dominant $N\,\pi\ \to N \,a$ process. 

Ref.~\cite{Coffey:2022eav} considered neutron star kinetic heating due to pseudoscalar-mediated DM interactions. 
These bounds can be seen in Fig.~\ref{fig:expinv} as a dark blue band at $f_a \sim 1\,\TeV$ for $m_\chi = 100\,\MeV$, and $10\,\GeV$.

Regarding direct detection, we have considered the strongest published spin-dependent bounds by CDMSlite~\cite{SuperCDMS:2017nns}, PICO-60L~\cite{PICO:2017tgi} and XENON-1T~\cite{XENON:2018voc} and 
 found that they are unable to exclude relevant regions of parameter space.
Spin-independent direct detection constraints can arise from a box diagram via the exchange of two axions (see Ref.~\cite{Arcadi:2017wqi}). 
However, the rate for such a process is proportional to $1/f_a^{8}$, and thus the constraints are expected to be further suppressed compared to the spin-dependent bounds we considered. We conclude that current direct detection limits do not play a role in constraining viable parameter space of our theory. 
 
Finally, the $a G \tilde G$ coupling may be generated by integrating out heavy quarks from a UV theory. 
In such a case, one would expect  that the heavy quarks appear below the scale $ 4\pi f_a$. Following Ref.~\cite{Hook:2019qoh}, a constraint can be placed of $4\pi f_a > 2\,\TeV$, with $2\,\TeV$  approximately the bound on new heavy quarks at the LHC~\cite{ATLAS:2019fgd,CMS:2018mts,ATLAS:2017jnp,CMS:2018mgb,ATLAS:2021mdj}. This model-dependent bound is illustrated by the light shaded gray regions in Figs.~\ref{fig:exp} and~\ref{fig:expinv}. Since it  depends on the UV completion of the theory, this constraint should not be considered as stringent as the other bounds we presented above.

\begin{figure}[t!]
	\centering	\includegraphics[width=0.48\textwidth]{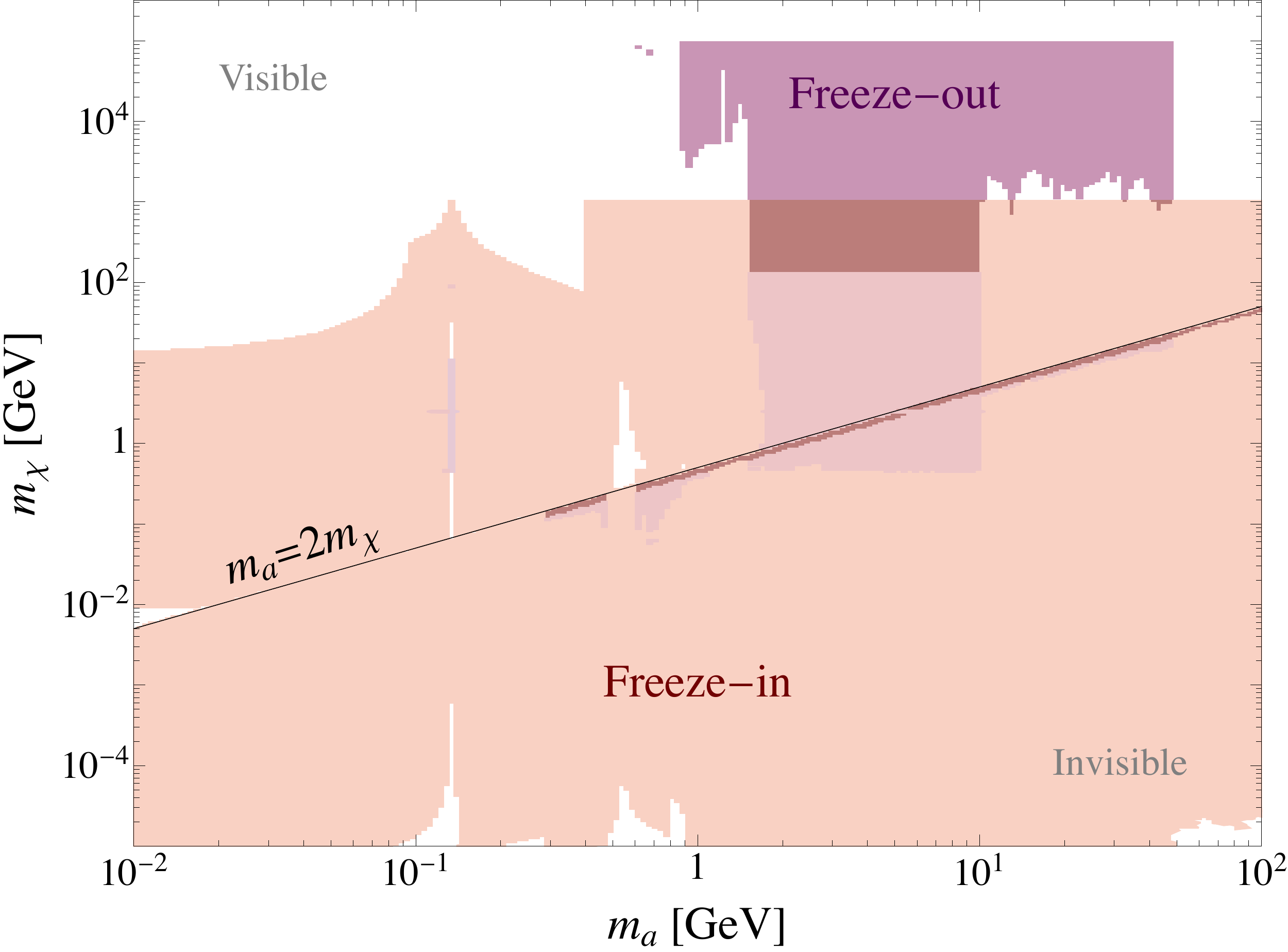}
	\caption{{\bf Allowed regions of DM-ALP parameter space.} A parameter scan consisting of $\sim 10^5$ points in the $m_a-m_\chi$ plane. For each set of masses we have numerically solved the Boltzmann equations and found a coupling $f_a$ such that the correct DM relic abundance is achieved. The colored regions represent points that have successfully solved the Boltzmann equations in the case of freeze out~(\textbf{{\color[rgb]{0.5, 0, 0.5}purple}})  and freeze in~(\textbf{{\color[rgb]{0.66, 0, 0}dark red}}) and are not ruled out by the existing limits as described in Section~\ref{sec:exp}. Freeze-out regions excluded only by the colored particles bound are shaded in \textbf{{\color[rgb]{0.8, 0.5, 0.8}light purple}}.
	\label{fig:expscan}
	}
\end{figure}

\section{Results}
\label{sec:results}

Our results for the axion-gloun coupling considered in this work are presented in Figs.~\ref{fig:exp}, \ref{fig:expinv} and \ref{fig:expscan}. Throughout we use $c_\chi=1$ and $\TRH=10\,\TeV$. 

Fig.~\ref{fig:exp} considers the visible decaying axion, where $m_a< 2m_\chi$. 
In the left panel of Fig.~\ref{fig:exp}, the colored shaded regions represent the constraints on the visibly decaying ALP. 
The solid lines delineate the freeze-in and freeze-out phases, while the dashed curves correspond to the numerical solutions to the Boltzmann equations for a fixed value of $m_\chi$.
In the right panel of Fig.~\ref{fig:exp}, the gray shaded regions represent the current constraints, the shaded regions outlined with solid curves indicate the freeze-in and freeze-out phases when including all relevant DM masses, such that the axion  decays visibly and our effective field theory description remains valid. Dashed curves show the projections for the reach of  future experiments.
We learn that for the visibly decaying axion, freeze-out is excluded by current experiments for axion masses $m_a$ between $10\,\MeV$ to a few hundred MeV.
The region between beam dumps and colliders at  $m_a\sim 1-50 \,\GeV$ remains viable for freeze-out, with decay constants of order $f_a\sim 100\,\GeV$ to $10\,\TeV$, and will be partially probed by Belle-II \cite{Chakraborty:2021wda}.
Freeze-in of the visibly decaying axion is currently constrained only by BBN and SN at axion masses $m_a \sim 10-400\, \MeV$, allowing for a broad range of axion masses with decay constant $f_a\gtrsim 100 \,\TeV$. The DUNE near detector~\cite{Kelly:2020dda}, CODEX-b~\cite{Gligorov:2017nwh}, LHC track trigger~\cite{Hook:2019qoh}, SHiP~\cite{SHiP:2015vad}, SHADOWS~\cite{Baldini:2021hfw}, NA62-LS3~\cite{NA62:2017rwk, Jerhot:2022chi} and MATHUSLA~\cite{Chou:2016lxi}  are expected to probe the freeze-in region further for ALP masses in the $m_a \sim 100\,\MeV - 20 \,\GeV$ range.

Fig.~\ref{fig:expinv} considers the invisible decaying axion, where $m_a > 2m_\chi$. Here we fixed the DM mass $m_\chi$ to several values, presented in the four panels of Fig.~\ref{fig:expinv}. 
For each value of $m_\chi$ we show the current constraints as colored shaded regions and the numerical solutions to the Boltzmann equations as dashed curves. 
 Note that since the DM mass is fixed, the $m_\chi = 0.1\,\GeV$ and  $10\,\GeV$ panels contain regions with visible axion decays, while for $m_\chi=100\,\GeV$ the figure is entirely visibly decaying axions. 
We learn that an invisibly decaying axion can avoid many of the constraints, allowing for a broad range of axion masses and decay constants.
In the freeze-out phase, near-resonance axion masses of $m_a \sim 2m_\chi$ enable significantly smaller couplings ({\it i.e.} larger decay constants $f_a$) than away from resonance. For $m_a\gtrsim 300\; {\rm MeV}$, this allows to evade existing limits, though a broader DM mass range is possible when the axion decays visibly. 
(Note that the freeze-out phase in the top left panel sits outside the plotted range and is excluded.) 
In the freeze-in phase, once invisible decays of the axion $a\to \chi\bar{\chi}$ become allowed, the coupling needed for freezing in the DM drops significantly compared to only visible decays; this is exemplified  by 
 the sharp change in couplings at $m_a = 2m_\chi$ in the upper right and bottom left panels of Fig.~\ref{fig:expinv}. 

Fig.~\ref{fig:expscan} summarizes a parameter scan in the $m_a-m_\chi$ plane consisting of $\sim 10^5$ points. 
For freeze-out, we considered $m_\chi > 10\,\MeV$ to avoid adding a thermalized relativistic degree of freedom during BBN, and $m_\chi< 10^3\,\TeV$ since values above this exceed the EFT validity region of $m_\chi < 8\pi f_a$.
For freeze-in, we focused on $m_\chi>10\,\keV$, corresponding to the freeze-in lower mass limit (see Ref.~\cite{DEramo:2020gpr}), and on $m_\chi< \TeV \ll \TRH$, in order to avoid solutions where $m_\chi$ is tuned close to $\TRH$ which would enable arbitrarily large couplings.

For each set of masses, we solved the Boltzmann equations numerically to find a coupling $1/f_a$ such that the correct relic abundance of DM is obtained. 
We then checked against existing constraints; the remaining allowed combinations of DM and axion masses are shown in the colored regions of Fig.~\ref{fig:expscan}: freeze-out~(purple) and freeze-in~(dark red). 
Freeze-out regions excluded only by the colored particles bound are shaded in light purple.
Note that some regions in this plane are able to accommodate the DM relic abundance both via freeze-out, with strong couplings such that $\chi$ and $a$ thermalize, and via freeze-in, through  smaller couplings where $\chi$ never thermalizes. 
The solid black curve corresponds to $m_a = 2m_\chi$, indicating the boundary between the visible (above) and invisibly (below) decaying~axion.

The allowed regions around axion masses $\sim 100\,\MeV$ and a few hundred MeV are narrow due to the proximity to the $\pi^0,\, \eta$ and $\eta'$ resonances which are difficult to probe. 
We find that freeze-out is viable for DM masses $m_\chi \gtrsim 30\,\MeV$ along with axion masses above $\sim 100\,\MeV$, when excluding the bounds on new colored particles, but is cut off around $m_\chi \simeq 100\,\GeV$ when including them.
The allowed DM-axion freeze-out mass region is more restricted in the case of invisibly decaying axions compared to the region for visible decays---most of the allowed parameter space exists  just below the $m_a = 2 m_\chi$ resonance. 
The freeze-in scenario is much less constrained than that of freeze-out due to the smaller couplings involved that are only partially probed by cosmological and astrophysical observations while remaining currently unprobed by terrestrial experiments.

\section{Summary}
\label{sec:conc}

In this work we considered the axion-gluon portal to dark matter in detail, considering different cosmological histories to explain the relic abundance. 
Studying both freeze-out and freeze-in processes, we have mapped out the cosmologically viable parameter space for DM and axion masses and couplings, along with the existing constraints from terrestrial experiments, cosmological considerations and astrophysical bounds. 

Future experiments will be able to probe the visibly decaying axion regions extensively. 
Belle-II \cite{Chakraborty:2021wda} is expected to improve current visible meson decay constraints; future runs of the LHC will statistically improve the current collider reach; and future experiments such as the DUNE near detector~\cite{Kelly:2020dda}, CODEX-b~\cite{Gligorov:2017nwh}, MATHUSLA~\cite{Chou:2016lxi}, FASER~\cite{Feng:2018pew,FASER:2018eoc}, SHiP~\cite{SHiP:2015vad}, SHADOWS~\cite{Baldini:2021hfw} and NA62-LS3~\cite{NA62:2017rwk, Jerhot:2022chi}
will extend the reach of beam dumps to a wider range of couplings in addition to masses $m_a \lesssim 10\,\GeV$ (see right panel of Fig.~\ref{fig:exp}). 
In addition, the proposed LHC displaced track trigger search is expected to probe a novel region of parameter space in the $m_a \sim 1-20\,\GeV$ range \cite{Hook:2019qoh, Kelly:2020dda}. 
Modification of the low energy supernova bounds~\cite{Caputo:2022mah} to include ALP-nucleon and ALP-pion couplings may also probe the currently open freeze-in region of the visibly decaying axion with $f_a \sim 10^5 -10^8\, \GeV$ at ALP masses $m_a < 1\; \GeV$. 
We conclude that the large coupling region $f_a \sim 100~\GeV- 10~\TeV$ at $m_a \sim 3-10\;\GeV$, which remains unconstrained by current and future experiments that we considered, is a well-motivated region for future terrestrial searches.

{\bf Acknowledgments.} We thank Martin Bauer, Jan Jerhot, and Kohsaku Tobioka for sharing of constraint data, and Edoardo Vitagliano for useful discussions related to stellar emission. 
P.F. is supported by the Zuckerman STEM Leadership Program.
The work of Y.H. is supported by the Israel Science Foundation (grant No. 1818/22), by the Binational Science Foundation (grant No. 2018140),  by the Azrieli Foundation and by an ERC STG grant (`Light-Dark', grant No. 101040019). 
E.K. is supported by the US-Israeli Binational Science Foundation (grant No. 2020220) and by the Israel Science Foundation (grant No. 1111/17). R.O. acknowledges support from the Israel Science Foundation (grant No. 1818/22).
Y.S. is supported by grants from the NSF-BSF (grant No. 2021800), the ISF (grant No. 482/20), the BSF (grant No. 2020300) and by the Azrieli foundation. 
This project has received funding from the European Research Council (ERC) under the European Union's Horizon Europe research and innovation programme (grant agreement No.\ 101040019).  
Views and opinions expressed are however those of the author(s) only and do not necessarily reflect those of the European Union. 
The European Union cannot be held responsible for them.

{\it Note Added:} While at final stages of writing this manuscript, we became aware of Refs.~\cite{Ghosh:2023tyz,Jeff:2023} that consider similar scenarios.
\appendix

\section{Rates}
\label{app:rates}

The general form for the Boltzmann equation for the number density of particle $X$, $n_X$, in a  Friedmann-Robertson-Walker  background is 
\begin{align}
    \frac{\partial n_X}{\partial t}+ 3H n_X= C_X \, ,  
\end{align}
where $C_X$ is the collision term, a sum of  all the rates of processes that can create or destroy an $X$ particle. 
In general for a $n\to m$ process 
\begin{align}
   \label{eq:process}
    i_1\, \cdots \, i_{n} \to  f_1 \, \cdots \,f_{m} \, ,
\end{align}
the rate is given by 
\begin{align}
    \gamma 
=   \Delta N_X \int d\Pi |\fM|^2 
    f_{i_1}(E_{i_1}) \ldots  f_{i_{n} }(E_{i_{n}}) \, ,
\end{align}
where $\Delta N_X$ is the number of $X$ particles created (or destroyed) in the process, $f_\ell(E)$ is the phase space density of a particle $\ell$, $|\fM|^2$ is the matrix element squared and summed over all degrees of freedom.
The $n+ m$ body phase space is given by
\begin{align}
    d\Pi 
=&  S\cdot d\Pi_{i_1} \cdots  d\Pi_{i_n}  
    d\Pi_{f_1} \cdots  d\Pi_{f_n} \times \nonumber\\
&   (2\pi)^4 \delta^{(4)}(\Sigma p_i-\Sigma p_f) \, ,
\end{align}
where $S$ is a symmetry factor if there are identical initial or final state particles and 
\begin{align}
    d\Pi_{\ell} = \frac{d^3 \vb{p}_\ell}{ (2\pi)^3 2 E_\ell} \, ,
\end{align}
is the Lorentz invariant phase space for particle $\ell$
with energy $E_\ell$. 
Here we have dropped additional contributions to the collision rates from quantum statistics, {\it i.e.}, Pauli blocking and stimulated emission.

Often one is interested in calculating collision terms for particles in thermal equilibrium. 
Ignoring the quantum statistics (which only have a small effect on the relic abundance calculations), the phase densities take on the familiar Maxwell-Boltzmann distribution 
\begin{align}
    f_\ell(E_\ell) 
=  \frac{n_\ell}{n_{\ell}^{\rm eq} } e^{-E_\ell/T_\ell} \, ,
\end{align}
where $n_\ell^{\rm eq}  = g_\ell \int  d\Pi_{\ell}  e^{-E_\ell/T_\ell}$. 
The collision rates can be written in terms of thermally averaged cross sections, which are defined as 
\begin{align}
    \label{eq:sigmavdef}
    \sigv_{i_1 \, \cdots i_{n} \to  f_1 \, \cdots \,f_{m}}&
    \equiv 
    \frac{|\Delta N_X|}{n^{\rm eq}_{i_1}\cdots n^{\rm eq}_{i_n}} \nonumber\\
    \times
    \int d\Pi & |\fM|^2  e^{-E_{i_1}/T_{i_1}}\ldots e^{-E_{i_n}/T_{i_n}} \, . 
\end{align}
Therefore, the collision rate 
is given by
\begin{align}
    \gamma_{ i_1 \,\cdots \, i_{n} \to  f_1 \, \cdots \,f_{m}}
=   n_{i_1}\cdots n_{i_n} \sigv_{i_1 \,\cdots\, i_{n} \to f_1 \,\cdots \,f_{m}}\, , 
\end{align}
where $n_i$ is the number density of particle $i$. 
For $2\to m$ processes, $\sigv$ turns out to be the cross-section times the Moeler velocity averaged over the phase space densities of the initial particles:
\begin{align}
    \sigv_{2\to m}
 =  \frac{1}{n^{\rm eq}_{i_1}n^{\rm eq}_{i_2} } 
    \int & d\Pi_{1} d\Pi_{2}  f_{i_1} f_{i_2}  \nonumber\\
    &\times \sigma\sqrt{(p_{i_1}\cdot p_{i_2})^2- m_{i_1}^2 m_{i_2}^2} \, .
\end{align}
For processes with more than two initial particles, we define $\sigv$ via Eq.~\eqref{eq:sigmavdef}.

For a $T$ (or $CP$) invariant process, the equilibrium rates for a process and its inverse are the same. 
Thus, the reverse rates are given by detailed balance as
\begin{equation}
    \sigv_{  f_1 \, \cdots f_{m} \to  i_1 \, \cdots i_{n} }^{\rm eq} 
    \!\!=\!\!
    \frac{n^{\rm eq}_{i_1} \cdots n^{\rm eq}_{i_n} }
    {n^{\rm eq}_{f_1} \cdots n^{\rm eq}_{f_m}}\!
    \sigv_{ i_1 \,\cdots i_{n} \to  f_1 \, \cdots f_{m}}^{\rm eq} \! .
\end{equation}
%

\subsection{Production through an intermediate axion}
\label{app:dmviaa}

In this appendix we estimate the DM production from the thermal bath, which may be dominated by on-shell and off-shell axions. 
For example, via processes such as $gq \to q (a^{(*)}  \to  \chi \bar{\chi})$ or  $gg \to (a^{(*)}  \to  \chi \bar{\chi})$. 
Since the matrix elements of these processes are factorized into axion production and axion decay, we can use the on-shell decay rates to 
calculate the off-shell axion production using the procedure described in this subsection.
We take the decay rate of the axion from Ref.~\cite{Aloni:2018vki}.
The DM production rate from the bath via  an intermediate axion, $X\to Y (a^*\to\chi\bar{\chi})$, 
is given by
\begin{align}
    \gamma_{X \to Y \chi \bar{\chi} } 
=&  \int  d\Pi_X d\Pi_Y d\Pi_{\chi_1}d\Pi_{\chi_2}
    (2\pi)^4 \delta^4(\Sigma_p p) f_X \nonumber\\
&  \times |\fM_{X \to Y \chi \bar{\chi} }|^2 \, ,
\end{align}
where $X$ and $Y$ represent some initial and final state of particles respectively; $d\Pi_X =d\Pi_{i_1} \cdots  d\Pi_{i_n}$ and $d\Pi_Y =d\Pi_{f_1} \cdots  d\Pi_{f_m}$ is the product of the Lorentz invariant phase space factors for the initial and final states 
and $f_X = f_{i_1 }(E_{i_1}) \cdots f_{i_n }(E_{i_n})$.
Factorizing the matrix element gives 
\begin{align}
    \gamma_{X \to Y \chi \bar{\chi} } 
=& \int  d\Pi_X d\Pi_Y d\Pi_{\chi_1}d\Pi_{\chi_2}
    (2\pi)^4 \delta^4(\Sigma_p p) f_X\no\\
&  \times\frac{ |\fM_{X \to Y a^*}|^2 |\fM_{a^* \to \chi \bar{\chi}}|^2}
    {(m_a^2-m_{a^*}^2)^2 - \Gamma_a^2 m_a^2} \, ,
\end{align}
where the matrix elements should be evaluated for an off-shell axion with $m_{a^*} = \sqrt{E^2_{a^*}-\vb{p}^2_{a^*}}\,$. 
Next, we can insert an identity integral over the internal axion 4-momentum
\begin{align}
    1 
=&   \int\frac{dm_{a^*}^2}{2\pi} d\Pi_{a^*} (2\pi)^4 \delta^4(p_{a^*}-p_X )\,.
\end{align}
Plugging this into the rate gives 
\begin{align}
    \label{eq:fulloffrate}
    \gamma_{X \to Y \chi \bar{\chi} }  
=&  \int \frac{dm_{a^*}^2}{\pi} \frac{\gamma_{X \to Y a^*} m_{a*}
    \Gamma_{a^* \to \chi \bar{\chi}} }{{(m_a^2-m_{a^*}^2)^2 - \Gamma_a^2 m_a^2}} \,, 
\end{align}
where we have used the definition of the production rate $ \gamma_{X \to Y a^*}$ and decay rate $\Gamma_{a^* \to \chi \bar{\chi}}$. 
These quantities should be evaluated for an off-shell axion with mass $m_{a^*}$. 
We can relate the rate $\gamma_{X \to Y a^*} $ to that found in Ref.~\cite{Aloni:2018vki,Salvio:2013iaa,DEramo:2021psx}. 
Using that the SM bath particles are always taken to be in equilibrium 
\begin{align}
    \gamma_{X \to Y a^*} 
=   n_{a^*}^{\rm eq } \Gamma_{a^*\,{\rm \sm}\to {\rm \sm}} \, .
\end{align}
As a sanity check, we take the narrow width approximation for the result in Eq.~\eqref{eq:fulloffrate} and find that $\gamma_{X \to Y \chi \bar{\chi} }  \to  \gamma_{X \to Y a} {\rm BR}\pqty{a\to \chi \bar{\chi}}$ as expected.

\subsection{\texorpdfstring{$2\to 2$}{} rates}

Following Ref.~\cite{Gondolo:1990dk}, the thermally averaged cross section from a $2 \to 2$ process $i_1 i_2\to f_1f_2$ is 
\begin{align}
    \label{eq:sigmavcmgen}
    \sigv
=&  \frac{T}{n_{i_1}^{\rm eq} n_{i_2}^{\rm eq}}
    \int \frac{ds\sqrt{s}}{512 \pi^5} K_1(\sqrt{s}/T) 
    \lambda^{\frac{1}{2}}(\sqrt{s},m_{i_1},m_{i_2}) \no \\
   \times &\lambda^{\frac{1}{2}}(\sqrt{s},m_{f_1},m_{f_2})
    \int \frac{d\Omega_{i_1,i_2}}{4\pi} 
    \frac{d\Omega_{f_1,f_2}}{4\pi}{|\mathcal{M}|^2} \, ,
\end{align}
where $d\Omega_{i,j}$ are taken in the $i,j$ center of mass frame, $K_i$ is the Bessel K function  and 
\beq 
\lambda(a,b,c)\equiv \left(1-(a+b)^2/c^2\right) \left(1-(a-b)/c^2\right)\,.\label{eq: lambda}
\eeq 
A useful limit is the $T\ll m$ limit, where 
\begin{align}    \label{eq:intsidentity}
 \frac{(2\pi)^3}{T^2e^{-2 m/T}}
    &\int ds\, K_1(\sqrt{s}/T) \left(1- \frac{4m^2}{s}\right)^{n}  \nonumber\\ 
    & \xrightarrow{T\to0}
   \  16\pi^{\frac{7}{2}} 
    \left(\frac{T}{m}\right)^{n-\frac{1}{2}} \Gamma(n+1)\, .
\end{align}

\subsection{\texorpdfstring{$\chi \bar{\chi} \rightarrow aa$}{}}
\label{app:chichitoaa}

The summed and squared matrix element for $\chi \bar{\chi} \rightarrow aa$ is
\begin{align}
    \label{eq:Mchichiaa}
    \abs{\mathcal{M}_{\chi \bar{\chi} \rightarrow aa}}^2 
=   &\frac{2c_\chi^4 m_\chi^4}{f_a^4} 
    \left[(m_\chi^2-t)(m_\chi^2-u)-m_a^4\right]\nonumber\\
    &\times\left(\frac{1}{t-m_\chi^2} -\frac{1}{u-m_\chi^2}\right)^2\,.    
\end{align}
A general closed form  expression for the thermally averaged cross section is difficult to find.

We use Eqs.~\eqref{eq:sigmavcmgen} and~\eqref{eq:Mchichiaa} to obtain
\begin{align}
    \sigv_{\chi \bar{\chi} \rightarrow aa}
=&   \frac{4 c_\chi^4 m_\chi^4}{512 \pi^5f_a^4}
    \frac{T}{(n_\chi^{\rm eq})^2 } \nonumber\\
    \times
&    \int ds  \sqrt{s}  K_1 (\sqrt{s}/T) 
    \left(\beta-\tan^{-1} \beta\right)\, ,
\end{align}
where $\beta\equiv\sqrt{1- {4m_\chi^2}/{s}}$ is the center of mass velocity of the $\chi$ and $\bar\chi$. 
In the limit that $T \ll m_\chi$, the velocity is small, $\beta \ll 1$, and the integral is dominated near $s=4m_\chi^2$. 
The thermally averaged cross section is then 
\begin{align}
    \sigv_{\chi \bar{\chi} \rightarrow aa}
=   \frac{ c_\chi^4 m_\chi^2}{1536 \pi^5 f_a^4}
    \frac{T}{(n_\chi^{\rm eq})^2 }
    \!\int\!\!ds   K_1 (\sqrt{s}/T) \beta^3  .
\end{align}
Using the limit in Eq.~\eqref{eq:intsidentity}, 
we find the thermally averaged cross section to be 
\beq
    \sigv_{\chi \bar{\chi} \to a a }
    \simeq \frac{c_\chi^4 m_\chi^2}{64\pi f_a^4} 
    \frac{T}{m_\chi}\,.
\eeq

\section{Freeze-in}
\label{app:freeze-in}

First, we consider DM production via $a$ decay, namely $a\to \chi \bar{\chi}$.
Ignoring the annihilation of $\chi$ paticles from inverse decays, the Boltzmann equation becomes
\begin{align}
    \label{eq:BoltzDM}
    \dot{n}_\chi + 3 H n_\chi 
=   2\left\langle \frac{m}{E}\right\rangle_a
    \Gamma_{a \to \chi \bar{\chi}} n_a  \, ,
\end{align}
where  
\begin{align}
    \label{eq: axion dilation factor}
    \left\langle \frac{m}{E}\right\rangle_a 
=   \frac{g_a}{n_a} \int\frac{ d^3 p }{(2\pi)^3} 
    \frac{m_a}{E_a} f_a
=   \frac{K_1(m_a/T)}{K_2(m_a/T)}\, ,
\end{align}
is the thermally averaged time dilation factor with $g_a=1$ the number of $a$ degrees of freedom.  In the last step we assume Maxwell-Boltzmann statistic. 

In terms of the DM and axion yields, where $Y_a=n_a/s$ and $Y_\chi=n_\chi/s$ with $s$ the entropy density, Eq.~\eqref{eq:BoltzDM} becomes
\begin{align}
    T \,  H \, \dot{Y}_\chi 
=   -2 \left\langle \frac{m}{E}\right\rangle_a 
    \Gamma_{a \to \chi \chi} Y_a \, .    
\end{align}
Therefore, the late time solution is given by 
\begin{align}
    Y_\chi(\infty) 
=   \int_0^{\TRH} dT \left\langle \frac{m}{E}\right\rangle_a  
    \frac{\Gamma_{a \to \chi \chi} Y_a }{T H}.
\end{align}

For the $a$ bath in equilibrium, that is $n_a =n_a^{\rm eq}$, this integral will be dominated in the IR near $T\sim m_a$. 
Assuming that $g_\star$ and $g_{\star s}$ are not changing rapidly near $T\sim m_a$, and that $m_a < \TRH$ an approximate solution can be obtained~\cite{Hall:2009bx}
\begin{align}
    Y_\chi(\infty) 
=   \frac{0.66 \ g_a}{g_{\star s}(m_a)\sqrt{g_{\star }(m_a)}} 
    \frac{m_{\rm pl}\Gamma_a}{m_a^2}\,.
\end{align}
Alternatively, it is possible that the $a$ abundance has frozen out, and has a constant abundance, $Y_a=$~constant, before it decays, then 
\begin{align}
Y_\chi(\infty) = 2 Y_a {\rm BR}(a\to \chi \bar{\chi}) \, . 
\end{align}
%

\section{Additional model details}
\label{app:model}

\subsection{Axion branching ratios and decay widths}

%
\subsubsection{\texorpdfstring{ $a \to \chi \bar{\chi}$}{}} 
\label{app: inv BR}

\begin{figure}[t!]
	\centering
	\includegraphics[width=0.48\textwidth]{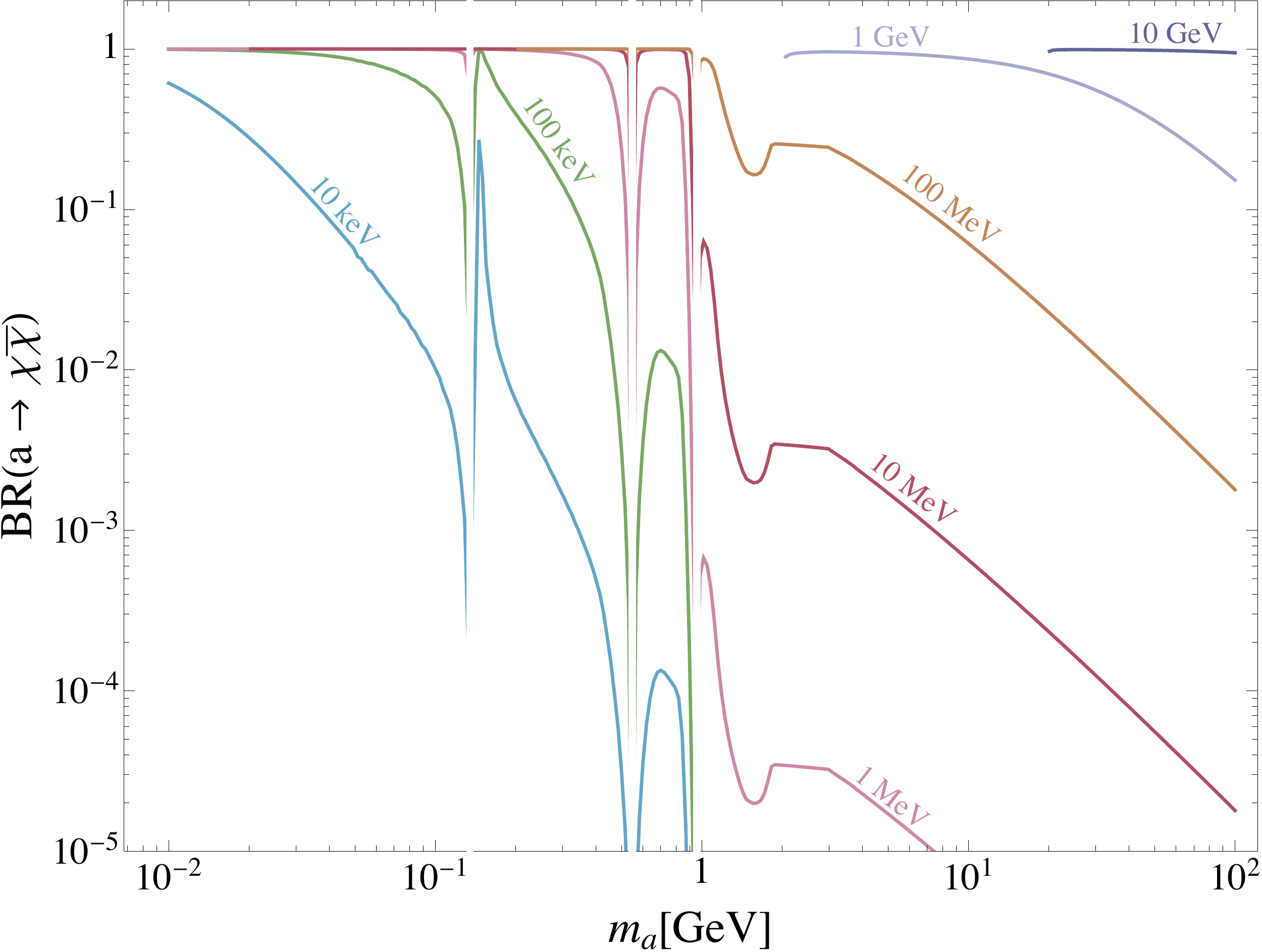}
	\caption{{\bf Branching ratio for the axion decay into DM.} Values of $\Br(a \to \chi \bar{\chi})$ for different DM masses $m\chi$, ranging from $10\,\keV$ to $10\,\GeV$, each depicted by a different colored curve.
    \label{fig: branching ratio to inv}
	}
\end{figure}

The branching ratio of the axion decay to DM for different values of $m_\chi$ is shown in Fig.~\ref{fig: branching ratio to inv}.
The non trivial features in the $m_a \sim 100\,\MeV - 2\,\GeV$ range mostly arise from the the $\pi^0$, $\eta$, and $\eta'$ resonances.
These features are evident in the freeze-in and freeze-out curves depicted in Figs.~\ref{fig:exp} and~ \ref{fig:expinv}.
The slight kink in the curves at $m_a = 3\,\GeV$ is due to a transition between two approximations of $\alpha_s$.
%

%
\subsubsection{\texorpdfstring{$a \to gg$}{}}

At $m_a \gtrsim 2\,\GeV$, where QCD becomes perturbative, the axion decay width to SM particles is given by it is decay width to gluons. 
This rate is given in Ref.~\cite{Spira:1995rr} to one-loop order
\begin{equation}
    \Gamma_{a \rightarrow gg} 
=   \frac{\alpha_s^2 m_a^3}{32\pi^3 f_a^2}\pqty{1+\frac{83\alpha_s}{4\pi}} \, .
\end{equation}

\subsubsection{\texorpdfstring{$a \to \gamma \gamma $}{}}

The axion-photon coupling is given by the coefficient $c_\gamma$ of the dimension-$5$ operator given in Eq.~\eqref{eq:aFFtil}.
Even when the bare axion-photon coupling vanishes at the UV scale $\Lambda$, two-loop contributions generate $c_\gamma$ at lower scales.  At $m_a \ll m_{\pi}$ the leading contribution comes from a chiral rotation of the $u,d,s$ quarks~\cite{Georgi:1986df, GrillidiCortona:2015jxo}
\begin{equation}
    c_\gamma \simeq 1.92 \pm 0.04 , \quad m_a \lesssim m_{\eta'}\,.
\end{equation}
The $u,d,s$ quarks have additional contributions at axion masses $m_a \lesssim 2.1\,\GeV$ originating from the the $a-P$ mixing and vector-meson photon mixing. 
At higher masses, $m_a \geq 2.1\,\GeV$ for $u,d,s$ and $m_a \geq 1.6\,\GeV$ for $c,b,t$, the running of quarks in the loop can be treated using perturbative QCD (pQCD). 
When the bare axion-quark couplings vanish this results in a contribution of order $\mathcal{O}\pqty{\alpha_s^2 \log(f_a)}$. 
A full quantitative discussion of the various terms contributing to $c_\gamma$ can be found in Ref.~\cite{Aloni:2018vki}, where the extension to masses $m_a \geq 3\,\GeV$ is found by replacing the pQCD contributions with the exact loop form factors found in Ref.~\cite{Bauer:2021mvw}.

Considering the effective axion photon interaction of Eq.~\eqref{eq:aFFtil}, the axion partial width into photons is given by 
\begin{align}
    \Gamma_{a\to \gamma \gamma} 
    = \frac{\alpha_{\rm EM}^2 m_a^3}{(8 \pi)^3 f_a^2} \abs{c_\gamma}^2 \, .
\end{align} 
%

\subsection{Meson decay rate to axion}

%
\subsubsection{\texorpdfstring{$V \to \gamma\, a$}{}}
\label{app: vector to gamma a}

Ref.~\cite{Bauer:2021mvw} provides a calculation of the branching ratio for the decay of quarkonium $V$, which is composed of quarks $q \bar{q}$, into a photon and an axion. 
The calculation takes into account one-loop radiative corrections and finds the expression
\begin{align}
    \label{eq: V to gamma a}
    \frac{\Br(V \to \gamma a)}{\Br(V \to e^+ e^-)} 
    & \approx 
    \frac{3 m_V^2 x Q_q^2}{8 \alpha_{\rm EM} f_a^2 (3\pi - \alpha_s)} \nonumber\\
    \times
&   \abs{c_{qq} \pqty{1-\frac{2 \alpha_s a_P(x)}{3\pi}} 
    - \frac{c_\gamma \alpha_s x}{\pi}}^2 ,
\end{align}
where $c_{qq}$ is axion-quark coupling at the scale $m_q$, $Q_q$ is the quark's electric charge, $x=1-m_a^2/m_V^2$ and $a_P(x)$ is a dimensionless monotonically increasing function of $x$, 
ranging with $a_P(0)=2$ and $a_P(1)\simeq6.62$, see Ref.~\cite{Bauer:2021mvw} for details.  
In our case, where $c_{qq}(\Lambda)=0$, $c_{qq}$ is generated from RG running from  $\Lambda$ to $m_q$. 
Following Ref.~\cite{Bauer:2021mvw}, we can write
\begin{align}
c_{cc}(m_c) \simeq -0.02\, , \qquad 
c_{bb}(m_b) \simeq -0.04\, ,
\end{align}
which correspond to running from $\Lambda = 4\pi\, \TeV$.
This approximation may lead to ${\cal O}(1)$ corrections to the bounds we have presented for $J/\psi \to \gamma\, a$ and $\Upsilon \to \gamma \, a$.

\subsubsection{\texorpdfstring{$B \to K\, a$}{}}
\label{app: B to K a running}

To recast the $B$ meson decays we use the result from Ref.~\cite{Chakraborty:2021wda}
\begin{align}
    \label{eq: B to K a}    
    \Gamma_{B \to K a}
=&  \frac{\left|C_W\right|^2 m_B^3}{64 \pi f_a^2}
    \left(1-\frac{m_K^2}{m_B^2}\right)^2 \lambda(m_B, m_K, m_a)^{1/2} \nonumber\\
&   \times\left[\frac{0.330}{1-m_a^2 / 37.5\,\GeV^2}\right]^2,
\end{align}
where $m_B\,(m_K)$ is the $B$-meson\,(kaon) mass and $C_W$ is a dimensionless constant multiplying the axion-bottom-strange vertex. 
A non-zero value of $C_W$ is induced by RG flow of the $aG\tilde{G}$ coupling from the UV scale $\Lambda$ to the electroweak scale (taken to be approximately the $W$-boson mass $m_W$).
Motivated by the analytical form of $C_W$, which was calculated to two-loop order in Ref.~\cite{Chakraborty:2021wda}, we approximate
for $\Lambda > m_W$
\begin{align}
    \label{eq:cW approx}
    C_W(\Lambda) 
=   & 0.2257 - 0.03428 \log \pqty{\frac{\Lambda}{\GeV}} \nonumber\\
    & - 0.0014 \log^2 \pqty{\frac{\Lambda}{\GeV}}\, ,
\end{align}
which holds to the 1\,\% level comparing to the full result in Fig.~ 3 of~\cite{Chakraborty:2021wda}.

In cases where $\Lambda < m_W$, for any of the above bounds, we discard it. 
This is justified as other constraints are typically stronger.

\subsection{\texorpdfstring{$a\bar{N}N$ and $a\bar{p}n\pi^+$ interactions}{}}

Following Ref.~\cite{DiLuzio:2020wdo,Chang:1993gm,Choi:2021ign,Lella:2022uwi} the axion-nucleon and axion-nucleon-pion interactions can be described by the effective Lagrangian
\beq
\begin{aligned} 
\label{eq:alp nucleon lagrangian}
\mathcal{L} \supset & \frac{\partial_\mu a}{2 m_N}\Big[g_{a p} \bar{p} \gamma^\mu \gamma_5 p+g_{a n} \bar{n} \gamma^\mu \gamma_5 n+  \\ & \quad +\frac{g_{ap} - g_{an}}{\sqrt{2} g_A f_\pi}\left(i \pi^{+} \bar{p} \gamma^\mu n-i \pi^{-} \bar{n} \gamma^\mu p\right)\Big]\, ,
\end{aligned}
\eeq
where $m_N$ is the nucleon mass and $g_A$ is a constant.
Ref.~\cite{Bauer:2021mvw} calculates the value of these coefficients 
\beq
\begin{aligned}
\label{eq:alp nucleon couplings}
g_{ap} = & \frac{m_N}{2f_a}\pqty{g_0 + g_A \delta_I \frac{m_{\pi_0}^2}{m_{\pi_0}^2 - m_a^2 + im_{\pi_0} \Gamma_{\pi_0}}},\\
g_{an} = & \frac{m_N}{2f_a}\pqty{g_0 - g_A \delta_I \frac{m_{\pi_0}^2}{m_{\pi_0}^2 - m_a^2 + im_{\pi_0} \Gamma_{\pi_0}}},
\end{aligned}
\eeq
where $\delta_I \equiv \frac{m_d - m_u}{m_d + m_u}$, $\Gamma_{\pi_0}$ is the $\pi_0$ decay width and with the values $g_A \simeq 1.25,\  g_0 \simeq 0.44$ taken from Ref.~\cite{Liang:2018pis}.

\section{Bounds from terrestrial experiments}
\label{app:casting constraints}

\subsection{Matching signal probabilities}
\label{app: matching signal probability}

We use the following simplified procedure to recast terrestrial constraints (meson decays, colliders and beam dumps), see \eg~\cite{Ilten:2018crw}.
We assume a measurement, which excludes a process $X \to Y\, (a \to f)$ with $X,Y,f$ denoting SM states and the on-shell ALP decay $a \to f$ occurring between times $\tau_1 < \tau_2$ in the axion rest frame. ($\tau_1,\tau_2$ are related to geometric lengths $L$ in the detector by $\tau = L/(\beta \gamma c)$, where $\beta,\gamma$ are the Lorentz transformation parameters and $c$ is the speed of light).
Assuming the measurement places a bound on the occurrence of more than $N$ such events, we find that $N$ factorizes into
\begin{align}
    \label{eq: number of events}
    N = N_X \times p(X \to Y \, a \mid X) \times p_{\rm detect} \, ,
\end{align}
where $N_X$ is the total number of $X$ states produced by the experiment (which is typically independent of the new physics), $p(X \to Y\, a \mid X)$ is the probability of the process $X \to Y\, a$ to occur given the initial state is $X$, and $p_{\rm detect}$ is the probability of detecting the final state $Y\, (a \to f)$ (which depends on $\tau_1,\tau_2$ and the branching ratio $\Br(a \to f)$) 
in addition to other experimental factors such as geometrical acceptance.
Typically, $p(X \to Y \, a \mid X) \propto f_a^{-2} \times \mathcal{O}({\rm polylog}(f_a))$ and in the case where $X$ is a single particle state $p(X \to Y \, a \mid X) = \Br(X \to Y\, a)$. 
We recast bounds by comparing the modifications introduced by our model to Eq.~\eqref{eq: number of events} with respect to the original analyses from which the bounds are obtained.

In cases where $f$ is a visible state, 
\begin{align}
    \label{eq: visible decay probability}
    p_{\rm detect}^{\rm visible} 
=   \Br(a\to f) \pqty{e^{-\frac{\tau_1}{\tau_a}} - e^{-\frac{\tau_2}{\tau_a}}} p_{\rm Eff}(Y\, f)\,,
\end{align}
where $\tau_a$ is the axion proper lifetime and $p_{\rm Eff}(Y\, f)$ is the detection efficiency of $Y\, f$ states which is assumed to encapsulate any additional experimental factors affecting the detection of $f$.
Throughout we assume $p_{\rm Eff}$ is independent of new physics.
 
In cases where $f$ is an invisible final state ({\it e.g.} appearing as missing energy), $\tau_2=\infty$ and the probability takes the form
\begin{align}
    \label{eq: invisible decay probability}
    p_{\rm detect}^{\rm invisible}  =  (1- B) + B \bqty{1- (1 - e^{-\frac{\delta}{B}}) p_{\rm Eff}} \, ,
\end{align}
where $B=\Br(a \to {\rm visible})$, $\delta =\Gamma_{a\to {\rm visible}} \tau_1$ (note that $\Gamma_a = \Gamma_{a \to {\rm visible}}/B$) and $p_{\rm Eff}$ is the efficiency of detecting the visible decay modes.
When recasting searches for invisible final states we place conservative bounds by taking $p_{\rm Eff}=1$ which ignores additional contributions from undetected axions decaying visibly within the decay volume.

Note that the addition of an invisible decay mode always results in a higher probability $p^{\rm invisible}_{\rm detect}$. 
The additional decay mode decreases $B$ while $\delta$ and $p_{\rm Eff}$ remain constant, thus, it is sufficient to show $p(B)$ is monotonically decreasing in $B$. 
We find
\begin{align}
    \pdv{p}{B} =  p_{\rm Eff} \bqty{-1 + \pqty{1 + \frac{\delta}{B}} e^{-\frac{\delta}{B}}} 
    <  0 \, ,
\end{align}
where we have used the positivity of $\delta$, $\sup_{x \in \mathbb{R}_+}\Bqty{(1 + x) e^{-x}} < 1$ and $0 < p_{\rm Eff} \leq 1$.

\begin{table*}[ht!]
    \centering 
    \label{tab: experimental paramters}
    \begin{tabular}{|l|c|c|c|c|c|}
        \hline
         Observable & References & $z_{\rm DV}$ & $\ell_{\rm DV}$ & RG flow \\ 
         \hline\hline
         $\Br(B^+ \rightarrow K^+ a(\mu^+ \mu^-))$& \cite{Bauer:2021mvw,LHCb:2016awg}~(LHCb) & $0$ & $0.74\,\text{m}$ & No \\ 
         \hline
         $\Br(K^+ \rightarrow \pi^+ a(\gamma\gamma))$& \cite{Bauer:2021mvw, E949:2005qiy}~(E949) & $0$  & $1.45\,\text{m}$ & No \\ 
         \hline
         $\Br(K^+ \rightarrow \pi^+ a(\gamma\gamma))$& \cite{Bauer:2021mvw, NA62:2014ybm}~(NA62) & $0$  & $140\,\text{m}$ & No \\ 
         \hline
         $\Br(K^+ \rightarrow \pi^+ a(\nu\bar{\nu})$& \cite{Bauer:2021mvw, NA62:2021zjw}~(NA62) & $140\,{\rm m}$ & $\infty$ & No \\ 
         \hline
         $\Br(B \rightarrow K a(3\pi))$ & \cite{Chakraborty:2021wda, Belle:2013nby}~(Belle) & 0 & $ 5\,\text{mm}$ & Yes \\ 
         \hline
         $\Br(B \rightarrow K a(\eta \pi \pi))$& \cite{Chakraborty:2021wda, BaBar:2011vod}~(BaBar) & 0 & $5\,\text{mm}$ & Yes \\ 
         \hline
         $\Br(B \rightarrow K a(K K \pi))$ & \cite{Chakraborty:2021wda, BaBar:2008rth}~(BaBar) & 0 & $5\,\text{mm}$ & Yes \\ 
         \hline
         $\Br(B \rightarrow K a(\phi\phi))$ & \cite{Chakraborty:2021wda,BaBar:2011vod}~(BaBar) & 0 & $5\,\text{mm}$ & Yes \\ 
         \hline
         $\Br(B \rightarrow K a(\gamma\gamma))$ & \cite{Chakraborty:2021wda,BaBar:2011vod}~(BaBar) & 0 & $5\,\text{mm}$ & Yes \\ 
         \hline
         $\Br(J/ \psi \to \gamma a(\gamma\gamma)$ & \cite{BESIII:2022rzz, AxionLimits}~(BESIII) & 0 & $\infty$ & No \\ 
         \hline
         $\Br(\Upsilon \to \gamma a({\rm hadrons}))$ & \cite{Bauer:2021mvw, BaBar:2011kau}~(BaBar) & 0 & 5\,{\rm mm} & No \\ 
         \hline
         $\Br({\rm \sm} \to {\rm \sm} ~ a(\gamma\gamma))$ & \cite{Jerhot:2022chi,Blumlein:1990ay}~(NuCal) & $64\,\rm m$ & $23\,\rm m$ & No \\ 
          \hline
          $\Br({\rm \sm} \to {\rm \sm} ~ a(\gamma\gamma))$ & \cite{Jerhot:2022chi,CHARM:1980zcj,CHARM:1983vkv,CHARM:1985anb}~(CHARM) & $480\,\rm m$ & $35\,\rm m$ & No \\
          \hline
          E137  & \cite{Dolan:2017osp,Bjorken:1988as}~(E137) & 179\,{\rm m} & 204\,{\rm m} & No \\ 
          \hline
          E141  & \cite{Dolan:2017osp,Riordan:1987aw}~(E141) & 12.16\,{\rm cm} & 35\,{\rm m} & No \\ 
          \hline
          LLP in EMD & \cite{Mitridate:2023tbj,CMS:2021juv}~(CMS) & $4\,\rm m$ & $3\,\rm m$ & No \\ 
          \hline
          $p\, p \to (a \to \gamma\, \gamma)$  & \cite{CMS:2014mvm, CMS:2015ocq, CMS:2017yta, ATLAS:2012fgo,ATLAS:2014jdv, ATLAS:2017cvh, ATLAS:2022abz, Mariotti:2017vtv, CMS-PAS-HIG-14-037, CMS-PAS-HIG-17-013, AxionLimits}~(CMS, ATLAS) & 0 & $1\, \rm{mm}$ & No \\
          \hline
          $p\, p \to (a \to j\, j)$  & \cite{ Mariotti:2017vtv, CMS:2017dcz}~(CMS) & 0 & $1\, \rm{mm}$ & No \\
          \hline
    \end{tabular}
    \caption{Summary of the parameters used to recast the terrestrial experiments. $z_{\rm DV}$ is the distance the axion must travel to the decay volume, $\ell_{\rm DV}$ is the length of the decay volume, RG flow indicates if RG-flow corrections to the couplings were~applied.}
    \label{tab: experimental parameters}
\end{table*}

\subsection{Approximating \texorpdfstring{$\tau_1, \tau_2$}{}}
\label{app: approximating t1 t2}

An experiment's dimensions allow us to relate $\tau_1$ to $\tau_2$. 
In particular, if we denote the distance the axion travels to the decay volume as $z_{\rm DV}$ and the length of the decay volume where the axion is detected as $\ell_{\rm DV}$ we find
\begin{align} 
    \label{eq: tau1 tau2 ratio}
    \frac{\tau_2 - \tau_1}{\tau_1} = \frac{\ell_{\rm DV}}{z_{\rm DV}} \, .
\end{align}
In the simple cases where $z_{\rm DV} = 0$ we take $\tau_1 = 0$ and where $\ell_{\rm DV} = \infty$ (as is the case for missing energy) we take $\tau_2 = \infty$. 
The values of $\ell_{\rm DV}, z_{\rm DV}$ we have used for each of the experiments considered in this work are consolidated in Table.~\ref{tab: experimental parameters}.

We use the following  approximations for $\tau_1$ and $\tau_2$ when relevant:
\begin{itemize}
    \item For a given $m_a$, when the bounds in the original analyses exclude $f_a$ in a certain range, $f_a^{\rm min} < f_a < f_{a}^{\rm max}$, we can numerically find $\tau_1,\tau_2$ by equating Eq.~\eqref{eq: number of events} for the upper and lower bounds $\eval{N}_{f_a^{\rm max}, m_a} = \eval{N}_{f_a^{\rm min}, m_a}$ and using Eq.~\eqref{eq: tau1 tau2 ratio} to find the bounds on our model.
    \item When the lab frame and the rest frame of the particle $X$ are approximately the
    same, we can approximate the boost of the axion.
    We ignore the angular distribution of the final states.
    In particular, for $X, Y$ that are single-particle states, the axion's boost and $\tau_1, \tau_2$ are given by
    \begin{align}
        \gamma \beta = \frac{m_X}{2 m_a} \lambda^{1/2}(m_X, m_Y, m_a)\, , 
    \end{align}
    and
    \begin{align}
        \tau_1 = \frac{z_{\rm DV}}{c \gamma \beta}\, ,\qquad 
        \tau_2 = \frac{z_{\rm DV}+\ell_{\rm DV}}{c \gamma \beta}\, ,
    \end{align}
    where $\lambda$ is defined in Eq.~\eqref{eq: lambda}.
\end{itemize}

\subsection{Meson decays}

Meson decay widths and their branching fractions may be modified in the presence of an axion-gluon coupling. 
Since we have taken all axion-SM couplings except the axion-gluon coupling to vanish at the UV scale $\Lambda = 8\pi f_a$ some of these effects are a result of RG running.

The bounds $B \to K\, (a \to 3\pi)$, $B \to K\, (a \to\eta \pi \pi)$, $B \to K (a \to KK \pi)$ and $B \to K\, (a \to \phi\phi)$ taken from Ref.~\cite{Chakraborty:2021wda} are corrected for RG running as is described above in Appendix \ref{app: B to K a running}. 
The bound $B^+\to K^+\, (a \to \mu^+ \mu^-)$ taken from Ref.~\cite{Bauer:2021mvw} is not corrected for RG running since it is already calculated by running from the approximately correct scale $\Lambda = 4\pi\,\TeV$. 
We have not taken into account the NLO effects of the RG flow in the various $K \to \pi\, a$ processes.
In quarkonia decays, $V \to \gamma\, a$ RG effects are not accounted for, and the branching ratios are calculated using the approximation described in Appendix \ref{app: vector to gamma a}.

When recasting the BaBar and Belle bounds we have assumed a prompt ALP decay corresponding to $z_{\rm DV}=0$ and $\ell_{\rm DV} = 5\,\text{mm}$ as was suggested in Ref.~\cite{Dror:2017nsg}. 
The BESIII analysis on $\Br(J/\psi \to \gamma \, a)$ in Ref.~\cite{BESIII:2022rzz} mentions only that the photon-coupled axion has a negligible decay width. 
Since any modifications our model introduces are only expected to make the ALP lifetime shorter, we place a conservative bound by ignoring the finite length of the BESIII detector decay volume.

\subsection{Beam Dumps}

We present constraints from di-photon measurements in proton beam-dumps from the NuCal~\cite{Blumlein:1990ay,Blumlein:2013cua} and CHARM~\cite{CHARM:1980zcj,CHARM:1983vkv,CHARM:1985anb} collaborations which are analyzed in Ref.~\cite{Jerhot:2022chi}. 
In addition, we show constraints from the electron beam dumps E137 \cite{Bjorken:1988as} and E141 \cite{Riordan:1987aw}, analyzed in Ref.~\cite{Afik:2023mhj}.
These bounds are  valid for the $m_a \leq 2m_\chi$ case where there are no invisible decays.
For  $2m_\chi < m_a$ where invisible decays are present, we recast these bounds by numerically finding $\tau_1$ and $\tau_2$ as  described in Appendix~\ref{app: approximating t1 t2}.
%

\subsection{Colliders}

We have used the LHC di-photon and di-jet bounds analyzed in Refs.~\cite{Mariotti:2017vtv, ATLAS:2022abz} based on the measurements performed by the CMS~\cite{CMS:2014mvm, CMS:2015ocq, CMS:2017yta, CMS:2017dcz, CMS:2019nrx, CMS-PAS-HIG-14-037, CMS-PAS-HIG-17-013} and ATLAS~\cite{ATLAS:2012fgo,ATLAS:2014jdv, ATLAS:2017cvh} collaborations.
In both cases, the original analysis assumed a GUT inspired model, with an ALP coupling to all three SM forces proportional to their coupling constants. 
We have approximated the di-jet production to be completely governed by the gluon coupling requiring only a rescaling of their gluon coupling. 
In di-photon searches, the narrow-width approximation is assumed where the cross-section factorizes to $\sigma(p\, p \to a) \Br(a\to \gamma\, \gamma)$.
We have approximated $\sigma \pqty{p\,p \to a}$ to depend only on the gluon coupling and cast the bounds by correcting for the significantly smaller $\Br(a \to \gamma\, \gamma)$ present in our model. 
In both cases, we considered prompt ALP decays with $z_{\rm DV} =0 $ and $\ell_{\rm DV} = 1\,{\rm mm}$.

Ref.~\cite{Mitridate:2023tbj} presented additional bounds on detection of long lived particles in the muon detection system of CMS~\cite{CMS:2021juv}.
We approximated the detector dimensions in this case as $z_{\rm DV} = 4\, {\rm m}, \, \ell_{\rm DV} = 3\,  {\rm m}$ ignoring any angular information, which may lead to an ${\cal O}(1)$ uncertainty on the bounds.

\section{Astrophysical and cosmological bounds}

\subsection{Supernovae}

We present the bounds from Ref.~\cite{Lella:2022uwi}, which consider the effects of axion-nucleon and axion-nucleon-pion couplings, see Eq.~\eqref{eq:alp nucleon lagrangian}, and axion-photon couplings, see Eq.~\eqref{eq:aFFtil}, on various observables.
Among them, we consider SN1987A cooling, ALP energy deposition in the mantle, non-observation of $\gamma$-rays from SN1987A, diffuse SN ALP background~(DSNALPB), and the expected $\gamma$-ray halo resulting from gravitational trapping of ALPs in Cassiopeia A. 
For the trapping regime (upper bound) of SN1987A cooling we use the estimation  $g_{ap} \leq 3 \times 10^{-9}$.
For the rest of the observables, we cast exclusions only for the displayed region in Fig.~4 of~\cite{Lella:2022uwi}.

There are two main assumptions used in the analysis that deviate from our model. 
The first assumption is that $g_{ap} \gg g_{an}$. 
This assumption breaks down in our model when $m_a \sim m_\pi$ where Eq.~\eqref{eq:alp nucleon couplings} dictates $g_{ap} \sim g_{an}$.
Since the majority of the axions are produced from $N\,\pi\ \to N \,a$, we have recast the bounds on $g_{ap}$ in Ref.~\cite{Lella:2022uwi} to bounds on $g_{ap} - g_{an}$ in our model Eq.~\eqref{eq:alp nucleon couplings}. 

The second assumption that requires altering is that the axion decays only to photons.
For the EFT considered in this work, the axion decays predominantly to other states at masses $m_a > \min \Bqty{3m_\pi, 2m_\chi}$.
To account for this we disregarded all bounds except SN1987A cooling for $m_a > \min\Bqty{3m_\pi,2m_\chi}$.
We find that the free-streaming regime (lower bound) of the SN1987A cooling bounds remains the same when the ALP decays invisibly while the trapping regime (upper limit) is expected to change significantly when a fraction of the axions decay to DM.
As the mean free path of DM in the SN core (that scales as $\sim f_a^{-4}$) is much larger than the axion's mean free path (that scales as $\sim f_a^{-2}$), we expect the trapping regime to extend to much larger couplings that are already excluded by terrestrial searches.
For simplicity, when $m_a > 2m_\chi$ we show exclusions for all couplings larger than those of the free-streaming bounds of SN1987A cooling, as can be seen in the upper panels of Fig.~\ref{fig:expinv}.
A dedicated analysis of low energy SN may add complementary bounds in the large coupling regime~\cite{Caputo:2022mah}.

\subsection{Neutron Star Heating}

The presence of an axion-gluon portal may have consequences that can be observed in stellar dynamics. 
Ref.~\cite{Coffey:2022eav} has placed bounds on such a model by considering the DM-induced kinetic heating of neutron stars. 
The bounds are presented only for the mass ratios $m_a/m_\chi= 1,\, 1/10$. 
As the $\chi-N$ cross-section is expected to be $m_a$ independent at $m_a \ll m_\chi$ we approximate the bounds at $m_a/m_\chi < 1/10$ to be the same as the bounds at $m_a/m_\chi = 1/10$. 
For $1/10 < m_a/m_\chi < 1$ we use a second-order polynomial interpolation to approximate the bounds.

\subsection{BBN}

The presence of a non-negligible abundance of long-lived axions during BBN may have measurable effects. 
Bounds considering the effects of an ALP with mass $m_a < 100\,\MeV$, which decays only to photons, have been placed 
in Refs.~\cite{Cadamuro:2011fd, Millea:2015qra, Depta:2020wmr, Kawasaki:2020qxm}.
We recast results from Ref.~\cite{Kawasaki:2020qxm}, which are agnostic to the production mechanism, to our model where the ALP is produced from other bath particles such as gluons. 
For such masses, the axion either freezes in at $T \sim \TRH$ to an abundance given by Eq.~\eqref{eq:axion freeze in} or freezes out when $\Gamma_{a\, \sm \to \sm} \sim H$ at temperatures $T \lesssim m_a$ which are larger than $T_{\rm BBN} \simeq 1 \, \MeV$. 
Recasting the results presented in Ref.~\cite{Kawasaki:2020qxm} is then straightforward, as each pair $\pqty{m_a,f_a}$ determines the axion lifetime $\tau_a$ and initial abundance $\eval{m_a Y_a}_{T=T_{\rm BBN}}$.

\newpage

\bibliographystyle{JHEP}
\bibliography{draft}

\end{document}